\documentclass{article}

\usepackage{arxiv}

\usepackage[utf8]{inputenc} 
\usepackage[T1]{fontenc}    
\usepackage{hyperref}       
\usepackage{url}            
\usepackage{booktabs}       
\usepackage{amsfonts}       
\usepackage{amsmath}
\usepackage{nicefrac}       
\usepackage{microtype}      
\usepackage{lipsum}
\usepackage{graphicx}
\graphicspath{ {./images/} }
\usepackage{comment}
\usepackage{cleveref}
\usepackage{natbib}
\usepackage{array}
\usepackage{multirow}
\usepackage{tikz}
\usepackage{float}
\usepackage{minted}
\usepackage{caption}
\usepackage{subcaption}
\usepackage{booktabs}
\usepackage{pdfpages}
\usepackage{makecell}
\usepackage{algorithm}
\usepackage{algpseudocode}
\usepackage{float}
\usepackage[section]{placeins}
\usepackage{nomencl}
\usepackage{lscape}

\graphicspath{{Fig/}}
\usepackage{tikz}
\newcommand{\tikzcircle}[2][red,fill=red]{\tikz[baseline=-0.5ex]\draw[#1,radius=#2] (0,0) circle ;}
\usepackage[locale=US,group-separator={,},separate-uncertainty=true,multi-part-units=single,per-mode=reciprocal]{siunitx}
\DeclareSIUnit\gCDW{g\textsubscript{CDW}}

\newcommand{\SuI}{SI}

\newcommand{\AC}{BMSA}
\newcommand{\ACfull}{Bayesian model set averaging}

\newcommand{\Ciso}{\textsuperscript{13}C}
\newcommand{\cmfa}{\texorpdfstring{\textsuperscript{13}C-MFA}{13C-MFA}}

\newcommand{\ECfull}{\textit{Escherichia~coli}}
\newcommand{\EC}{\textit{E.~coli}}
\newcommand{\tri}{\emph{Triangulus}}

\newcommand{\netflux}[1]{$v^n_{#1}$}
\newcommand{\xchflux}[1]{$v^x_{#1}$}
\newcommand{\tracerA}{\SI{80}{\percent}~[U-\textsuperscript{13}C] + \SI{20}{\percent}~[\textsuperscript{12}C]}
\newcommand{\tracerBa}{\SI{100}{\percent}~[1,2-\textsuperscript{13}C]}
\newcommand{\tracerBb}{\SI{100}{\percent}~[1,6-\textsuperscript{13}C]}
\newcommand{\tracerBc}{\SI{50}{\percent}~[1-\textsuperscript{13}C] + \SI{50}{\percent}~[4,5,6-\textsuperscript{13}C]}
\newcommand{\InlineCmt}[1]{\text{\algorithmiccomment{#1}}}

\crefname{section}{sec.}{sec.}
\Crefname{section}{Sec.}{Sec.}
\crefname{figure}{fig.}{fig.}
\Crefname{figure}{Fig.}{Fig.}

\usepackage{relsize}
\def\ifmonospace{\ifdim\fontdimen3\font=0pt }

\def\C++{%
\texorpdfstring{
\ifmonospace%
    C++%
\else%
    C\kern-.1067em\raise.35ex\hbox{\smaller[2]{++}}%
\fi%
\spacefactor1000}{C++} }

\makenomenclature

\title{Trans-dimensional Bayesian model averaging for \textsuperscript{13}C-based metabolic flux analysis: Evidence-based flux inference under structural model uncertainty}

\author{
 Johann F. Jadebeck\textsuperscript{1,2}, Anton Stratmann\textsuperscript{1,2}, Martin Beyß\textsuperscript{1,2}, Katharina N{\"o}h\textsuperscript{1,$\ast$} \\[2ex]
  \textsuperscript{1}Institute of Bio- and Geosciences, IBG-1: Biotechnology,
  Forschungszentrum Jülich, Jülich, Germany \\
  \textsuperscript{2}Computational Systems Biotechnology (AVT.CSB),
  RWTH Aachen University, Aachen, Germany \\[2ex]
  \textsuperscript{$\ast$} {Corresponding author \href{email:k.noeh@fz-juelich.de}{k.noeh@fz-juelich.de}}
}

\begin{document}

\maketitle

\begin{abstract}
Accurate quantification of intracellular metabolic fluxes is central to systems biology and biotechnology. Flux estimation relies on biochemical network models, with \textsuperscript{13}C metabolic flux analysis (MFA) being the state-of-the-art approach. However, isotope labeling data are often insufficient to uniquely support a single network formulation. In such cases, flux estimates become model-dependent, highlighting the need for methods that explicitly account for structural uncertainty. Bayesian model averaging (BMA) provides a principled framework for this purpose, but its application to \textsuperscript{13}C-MFA has so far been restricted to uncertainty in reaction bidirectionality within fixed network topologies.
We introduce a scalable Bayesian inference framework for \textsuperscript{13}C-MFA, \ACfull{} (\AC), that applies BMA to encompass uncertainty in reactions and pathways. Our approach combines reversible jump Markov chain Monte Carlo for trans-dimensional exploration of model spaces with diffusive nested sampling for robust estimation of model evidences, enabling averaging over large families of metabolic network models. Using illustrative and application-scale synthetic case studies, we demonstrate that the method yields robust flux estimates, reveals when multiple network configurations are statistically indistinguishable, and recovers data-supported model structures. 
Importantly, rather than committing to a single model, the framework manages structural uncertainty: under limited data, competing models are retained, whereas increasing data informativeness improved model and flux recovery. The approach scales to billions of model variants, providing a practical foundation for uncertainty- and misspecification-aware quantitative flux inference in \textsuperscript{13}C-MFA.
\end{abstract}

\keywords{\textsuperscript{13}C metabolic flux analysis, Bayesian model averaging, Diffusive nested sampling, Reversible jump MCMC, Model uncertainty, Isotope labeling data}

\section{Introduction}
\label{sec:introduction}
Metabolism plays a central role in systems biology, biotechnology, and metabolic engineering. Quantitative understanding of intracellular metabolic processes is essential for the design of sustainable bioprocesses~\citep{Liu2025}, for dissecting host-pathogen interactions and drug responses~\citep{Beste2013}, and for uncovering metabolic reprogramming in cancer~\citep{Antoniewicz2018}. Decades of biochemical research and genomics, complemented by computational platforms~\citep{Gong2024}, have yielded comprehensive reconstructions of the biochemical reactions that \textit{can} occur in a cell. However, determining which of these reactions are active under a given physiological state, and at what rates, requires condition-specific data. Reaction activity depends strongly on physiological state and environmental context: for example, latent metabolic pathways may only become active under genetic or environmental perturbations~\citep{Fong2006}.

Metabolic fluxes, represent the integrated functional outcome of cellular processes and are therefore central to biological interpretation~\citep{Sauer2006}. However, fluxes are not directly observable and must be inferred from data using a mechanistic models. Isotope-based metabolic flux analysis (\cmfa) is a widely used technology for estimating intracellular fluxes~\citep{Niedenfuhr2015,Zamboni2009,Long2019}. It combines isotope-labeling patterns and extracellular exchange rates collected under metabolic (pseudo)steady-state conditions with a metabolic model to solve an inverse problem. This inverse problem is typically ill-posed, such that not all fluxes can be determined with high confidence~\citep{Wiechert2021}. Advances in the automation of isotope-labeling experiments (ILE) and the integrative analysis of multiple datasets using different tracers improve precision~\citep{Leighty2013}. Nevertheless, many fluxes remain non-identifiable due to insufficiently data informativeness andstructural non-identifiability inherent to metabolic cycles~\citep{Kappelmann2016}.

A fundamental challenge in \cmfa{} is its dependence on the assumed metabolic network model. The model must capture the relevant metabolic pathways active under the studied experimental condition~\citep{Long2019,Theorell2024}. Simply including all known reactions is not a solution: large networks introduce alternative routes that cannot be discriminated by the available data, increasing non-identifiability and the risk of biologically implausible solutions~\citep{Linden-Santangeli2025}. For this reason, curated core models are commonly used to maintain identifiability. However, structural uncertainty remains even within such models. In particular, the presence of latent pathways can lead to substantially different flux estimates depending on the chosen model formulation ~\citep{Fong2006,Zamboni2009}, underlining that flux estimation is inherently model-dependent.

In practice, this dependency gives rise to a model selection problem: identifying which model(s) are supported by the data. Standard approaches based on goodness-of-fit are insufficient for reliable model discrimination~\citep{Sundqvist2022}. Information criteria such as AIC or BIC account for model complexity but become difficult to apply in combinatorial model spaces~\citep{Theorell2024}. Validation-based approaches using held-out data~\cite{Sundqvist2022} provide an alternative, but require sufficiently comprehensive datasets, limiting their capability when data are scarce or redundant.

We here present an alternative approach that avoids committing to a single model. We adopt a Bayesian framework to explicitly account for structural model uncertainty at the level of reactions and pathways. Using Bayesian model averaging (BMA, \cite{Hoeting1999}), we treat network structure (the set of reactions that constitutes the model) as a variable and propagate both data and structural uncertainty into flux estimates by averaging over a family of candidate models. The candidate model set is constructed from biochemical knowledge and assumed to be sufficiently comprehensive to capture the underlying metabolic processes.

Instead of conditioning inference on a single model, BMA yields flux estimates that reflect the support of all and possibly competing model hypotheses. By weighting models according to their posterior probability, the approach quantifies relative support within the model space and reduces overconfidence associated with single-model inference. Notably, posterior model probabilities are interpreted as evidence given the data rather than as claims of model correctness. 

In this work, we generalize the approach of \citet{Theorell2024}, who considered uncertainty in reaction bidirectionality within a fixed network, to also account for uncertainty in reactions and pathways. The resulting inference problem is computationally challenging, as it involves nonlinear, potentially multimodal likelihoods defined over linearly constrained flux spaces that vary across models. 
To address this, we develop a scalable inference algorithm that combines reversible jump Markov chain Monte Carlo (RJMCMC) for trans-dimensional exploration of the model space with diffusive nested sampling for robust and interpretable model evidence estimation~\citep{Green1995,Brewer2011}. 
We evaluate our approach using simulated data, focusing on flux inference under structural uncertainty for both illustrative and application-scale \cmfa{} scenarios. This allows us to systematically analyze how the BMA approach provides uncertainty-aware inferences across real-world regimes of data informativeness.

\section{Problem statement}
\label{sec:problem}
We first revisit conventional single-model flux inference and an existing multi-model approach that accounts for uncertain reaction bidirectionality. We then generalize the problem to include uncertainty at the level of reactions and pathways.

\subsection{Bayesian single-model metabolic flux inference}

Given an ILE dataset $\mathcal{D}$, the conventional Bayesian approach uses a single network model $\mathcal{M}$ and infers the posterior probability density $p(v_\mathcal{M} \mid \mathcal{D})$ of the model-specific fluxes $v_\mathcal{M}$~\citep{Theorell2017}. The model $\mathcal{M}$ consists of a set of relevant biochemical reactions, which we refer to as \textit{reaction set}, their associated atom transitions, and their assumed bidirectionality~\citep{Wiechert1997}. Note that bidirectional reactions introduce both net and exchange fluxes, whereas unidirectional reactions contribute only net fluxes, while their exchange fluxes are zero. 

Under metabolic (quasi)steady-state conditions, the reaction stoichiometry of $\mathcal{M}$ together with physiological flux limits induces linear mass balance constraints that restrict fluxes to a bounded, convex polytope $\mathcal{P}_\mathcal{M}$~\citep{Theorell2022}. The dimensionality of this feasible flux space defines the degree of freedom (DOF) of the inference problem. The flux posterior $p(v_\mathcal{M} \mid \mathcal{D})$ is typically explored using Markov chain Monte Carlo (MCMC)~\citep{Theorell2017}.

\subsection{Flux inference at uncertain reaction bidirectionality}

A key limitation of single-model \cmfa{} is the assumption that the model structure is correct. In particular, reaction bidirectionality depends on \textit{in vivo} thermodynamic driving forces that are not precisely known~\citep{Wiechert2007}, and exchange fluxes are typically weakly identifiable~\citep{Wiechert1997bidir2}. This combination gives rise to a large number of model variants that differ in reaction bidirectionality and, consequently, associated DOFs.
Specifically, for $n_x$ reactions with uncertain bidirectionality, $2^{n_x}$ unique model variants exist. For typical \cmfa{} models, the number of model variants ranges from a few dozen to several hundred thousand~\cite{Theorell2024}.

To account for this uncertainty, the finite set $\{ \mathcal{M} \}_{\mathcal{I}}$ of candidate models is considered and BMA is applied, which combines the single-model flux posteriors $p( v_{\mathcal{M}_i} \mid \mathcal{D})$ with the associated posterior model probabilities $p( \mathcal{M}_i \mid \{ \mathcal{M} \}_{\mathcal{I}}, \mathcal{D})$ conditioned on the model set in a weighted average to determine model set-averaged fluxes $v_{ \{ \mathcal{M} \}_{\mathcal{I}}}$ 
\begin{equation}
    p( v_{ \{ \mathcal{M} \}_{\mathcal{I}}} \mid \mathcal{D}) 
    = \sum_{i' \in \mathcal{I}} 
        p( \mathcal{M}_{i'} \mid \{ \mathcal{M} \}_{\mathcal{I}}, \mathcal{D}) 
        \cdot p( v_{\mathcal{M}_{i'}} \mid 
        \mathcal{D})
    \label{eq:BMA}
\end{equation}
where model set-averaged fluxes are defined on the shared flux space $\mathcal{P}_{ \{ \mathcal{M} \}_{\mathcal{I}}}$, by assigning zero values to parameters absent in a given model. Here, the single-model probabilities $p(\mathcal{M}_i \mid \{ \mathcal{M} \}_{\mathcal{I}},\mathcal{D})$ act as weights representing the relative support of each variant given $\mathcal{D}$, thereby encoding the model's ability to describe the data.

Direct evaluation of all posterior model probability in Eq.~\eqref{eq:BMA} is computationally prohibitive, as it requires high-dimensional integration over convex polytopes. RJMCMC provides a practical alternative by jointly sampling (continous) flux and (discrete) model spaces, thereby avoiding explicit enumeration of millions of model evidences~\citep{Green1995, Theorell2020}. This approach has been shown to improve the robustness of flux inference~\citep{Theorell2024} and is applied in \cmfa{}~\cite{BorahSlater2023}.

\subsection{Flux inference at full model uncertainty}

While the approach by \citet{Theorell2020} accounts for uncertainty in reaction bidirectionality, it still assumes that the underlying reaction set is correct. This assumption is often questionable, particularly in the presence of latent pathways whose activity, i.e. whether a pathway carries a non-zero net flux, is condition-dependent~\citep{Zamboni2009}.

We therefore extend the formulation in Eq.~\eqref{eq:BMA} to account for uncertainty at the level of both reaction sets (reactions and pathways) and reaction bidirectionality. 
Let $\mathcal{M}^k$, $k = 1 \ldots K$ denote $K$ alternative reaction sets, each containing $n_{x,k}$ reactions with uncertain bidirectionality. Then each reaction set induces a model set $\{ \mathcal{M}^k \}_{\mathcal{I}^k}$ with $2^{n_{x,k}}$ variants. 
The full candidate model space is then given by 
\begin{equation}
    \{\mathcal{M}^k\}_{\mathcal{K}} =
    \bigcup\limits_{k'=1}^{K} { \{ \mathcal{M}^{k'} \}_{\mathcal{I}^{k'}}}
    \quad\text{with}\quad \mathcal{K} = \bigcup\limits_{k'=1}^{K} \mathcal{I}^{k'}
    \label{eq:fullmodelset}
\end{equation}
containing $\sum_{k'=1}^{K} 2^{n_{x,k'}}$ model variants.

The BMA formulation extends naturally to this setting: the model set-averaged flux posterior given the full model set is 
\begin{equation} 
    p( v_{\{ \mathcal{M}^k \}_\mathcal{K}} \mid \mathcal{D}) = 
    \sum\limits_{k' = 1, \, i' \in \mathcal{I}^{k'}}^K
        p( \mathcal{M}^{k'}_{i'} \mid \{ \mathcal{M}^{k'} \}_{\mathcal{K}}, \mathcal{D} ) 
    \cdot  p( v_{\mathcal{M}^{k'}_{i'}} \mid \mathcal{D})
    \label{eq:full-BMA} 
\end{equation}
with the single-model posterior probabilities for any candidate model $\mathcal{M}^{k'}_{i'}$ in view of the full model set
is given by
\begin{equation}
    p( \mathcal{M}^{k'}_{i'} \mid \{ \mathcal{M}^{k'} \}_{\mathcal{K}}, \mathcal{D}) = \\
    \frac{p ( \mathcal{D} \mid \mathcal{M}^{k'}_{i'}) \cdot p(\mathcal{M}^{k'}_{i'})}
    {\sum\limits_{k''=1, i'' \in \mathcal{I}^{k''}}^{K} \ p ( \mathcal{D} \mid \mathcal{M}^{k''}_{i''} ) \cdot p ( \mathcal{M}^{k''}_{i''})}
    \label{eq:modelprob}
\end{equation}
with model priors $p(\mathcal{M}^{k}_{i})$. Herein, a key quantity is the model evidence 
\begin{equation}
    p ( \mathcal{D} \mid \mathcal{M}^k_i ) = \int_{\mathcal{P}_{\mathcal{M}^k_i}} p( \mathcal{D} \mid v_{\mathcal{M}^k_i}) \cdot p ( v_{\mathcal{M}^k_i} \mid \mathcal{M}^k_i ) \ d v
    \label{eq:evidence}
\end{equation}
with the model-specific likelihood $p( \mathcal{D} \mid v_{\mathcal{M}^k_i})$ and flux priors $p ( v_{\mathcal{M}^k_i} \mid \mathcal{M}^k_i )$. Evaluating these quantities across combinatorial model spaces constitutes the main computational challenge we address in this work.

\subsection{A note on priors in \cmfa}
\label{ssec:priors}

The Bayesian formulation in Eqs.~\eqref{eq:modelprob}-\eqref{eq:evidence} requires priors over both model structures and flux parameters. As in all Bayesian model selection frameworks, model probabilities and evidences may be sensitive to prior choices~\citep{kass1995bayes}, making careful prior specification an important aspect of practical applications.
Candidate reaction sets are typically derived from biochemical knowledge bases such as KEGG (\url{https://www.genome.jp/kegg/}) and BioCyc (\url{http://biocyc.org/}), but these resources do not resolve condition-specific reaction or pathway activity, nor reaction bidirectionality~\citep{Fong2006,Nishikawa2008}. 

In this work, we adopt the following pragmatic choices. 
First, we assume that the candidate model space is sufficiently comprehensive to capture the relevant metabolic processes, while excluding implausible variants (near $\mathcal{M}$-closed setting). Consequently, posterior model probabilities are interpreted as relative support within this space.
Second, to avoid bias toward reaction sets that induce more model variants (due to a larger number of reactions with uncertain bidirectionality), we assign equal prior mass to each reaction set and uniform model priors within each set, yielding $p(\mathcal{M}_i^k \mid \{ \mathcal{M}^k \}_{\mathcal{K}}) = 1/(K \cdot 2^{n_{x,k}})$. 
Third, for the single-model fluxes we use uniform priors over the respective flux polytope $\mathcal{P}_{\mathcal{M}_i^k}$, which are proper by construction and encode biochemical constraints, thereby providing a weakly informative default choice~\citep{Theorell2024}.
More generally, prior information on reaction/pathway activity, reaction bidirectionality, or flux ranges can be incorporated when available. 

\section{A divide-and-conquer approach to flux inference}
\label{sec:ModelJumping}

Directly applying RJMCMC to infer model-averaged fluxes under full model uncertainty is challenging. Beyond the combinatorial size of the model space, the sampler must explore the associated continuous, model-specific polytope-constrained flux spaces. In particular, the polytopes impose complex constraints to the net flux coordinates~\citep{Jadebeck2023}, making the design of efficient trans-dimensional jump proposals difficult. As a result, naive RJMCMC implementations suffer from low acceptance rates and slow convergence. 
To address this, we adopt a divide-and-conquer strategy that breaks down the inference problem into tractable sub-problems on the level of reaction set, while tackling the full problem complexity in a computationally efficient manner.

\subsection{Dividing flux inference into smaller problems}

We partition the full model set $\{\mathcal{M}^k\}_\mathcal{K}$ into $K$ disjoint subsets $\{\mathcal{M}^k\}_{\mathcal{I}^k}$, each corresponding to a fixed reaction set. Within each model subset, uncertainty is restricted to reaction bidirectionality, allowing the use of established RJMCMC methods~\citep{Theorell2020}. 

Expressed in terms of these subsets, the full model set-averaged flux posterior in Eq.~\eqref{eq:full-BMA} is written as (see \SuI{}~Appendix~\ref{SI:S1_BMA_modelsets}) 
\begin{equation}
    \begin{aligned}
       p( v_{\{ \mathcal{M}^k \}_\mathcal{K}} \mid \mathcal{D}) =  
       \sum\limits_{k'=1}^{K} \
        p( \{\mathcal{M}^{k'}\}_{\mathcal{I}^{k'}} \mid 
        \mathcal{D}) 
        \cdot 
        p( v_{\{\mathcal{M}^{k}\}_{\mathcal{I}^{k'}}} \mid \mathcal{D})
        \label{eq:BMA-with-model-sets}
    \end{aligned}
\end{equation}
i.e., as a weighted combination of model subset-averaged flux posteriors.
The corresponding posterior probabilities of the model subsets are given by
\begin{equation}
    \begin{aligned}
        p( \{\mathcal{M}^{k'}\}_{\mathcal{I}^k} \mid \mathcal{D}) =
        \frac{
        p(\mathcal{D} \mid \{\mathcal{M}^k\}_{\mathcal{I}^k}) \cdot
        p(\{\mathcal{M}^k\}_{\mathcal{I}^k})
        }{\sum\limits_{k'=1}^{K}\
        p(\mathcal{D} \mid \{\mathcal{M}^{k'}\}_{\mathcal{I}^{k'}})\cdot p(\{\mathcal{M}^{k'}\}_{\mathcal{I}^{k'}})}
    \end{aligned}
    \label{eq:model-subset-probability}
\end{equation}
The key quantity herein is the evidence of each model subset, which is obtained by aggregating the evidences of all single models within the subset
\begin{equation}
    p(\mathcal{D} \mid \{\mathcal{M}^k\}_{\mathcal{I}^k}) =
    \sum\limits_{i'\in\mathcal{I}^k}
    p(\mathcal{M}_{i'}^k \mid \{\mathcal{M}^k\}_{\mathcal{I}^k}) 
    \cdot p (\mathcal{D} \mid \mathcal{M}^k_{i'})
    \label{eq:model-subset-evidences}
\end{equation}
This reformulation transforms the original inference problem in Eqs.~\eqref{eq:full-BMA}-\eqref{eq:evidence}, into a finite collection of $K$ smaller BMA problems, each defined on a fixed reaction set and therefore amenable to RJMCMC-based inference~\citep{Theorell2020}. 

\subsection{Conquering subset-level problems}

To obtain the model set-averaged flux posterior in Eq.~\eqref{eq:BMA-with-model-sets}, accurate estimation of the evidences $p ( \mathcal{D} \mid \{\mathcal{M}^k\}_{\mathcal{I}^k})$ is required. For this, we employ diffusive nested sampling (DNS), a flexible framework for evidence estimation in multimodal and trans-dimensional settings~\citep{Brewer2011, Brewer2015_inference}.
DNS is an iterative algorithm that uses MCMC to sample from regions of increasing likelihood. When combined with RJMCMC as the exploration mechanism, it enables sampling across both discrete model and continuous parameter spaces~\cite{Brewer2015_lens}. In our case, these correspond to  model structure variants within a model subset and their associated flux polytopes. 
Embedding RJMCMC within DNS is known as trans-dimensional DNS (TDNS)~\citep{Brewer2015_inference, Brewer2015_lens}.
Our key point is that, by restricting trans-dimensional moves to variations in reaction bidirectionality, TDNS is applicable directly to the subset-level problems. This allows us to reuse established RJMCMC proposals~\citep{Theorell2020} without modification and avoids the need to design new cross-model proposals across reaction sets, which would otherwise be prohibitively complex.

The resulting subset evidences are used to compute posterior model subset probabilities in Eq.~\eqref{eq:model-subset-probability}), posterior single-model probabilities (\SuI{}~Appendix~\ref{SI:S2_TDNS}), and, by combining subset-level posteriors according to Eq.~\eqref{eq:BMA-with-model-sets}, the full model set-averaged flux posterior. We refer to this algorithm as \textit{\ACfull{}} (\AC), with pseudocode given in Algorithm~\ref{alg:BMA}. The dominant computational cost arises from the subset-level inference and evidence estimation (L~12), while the final aggregation across the $K$ subsets is negligible (L~14,16,18); cf.~\SuI{}~Appendix~\ref{SI:S2_TDNS} for details.

\begin{algorithm}
\caption{\AC: \cmfa{} under model uncertainty}
\label{alg:BMA}
    \begin{algorithmic}[1]
    \State \textbf{input:} 
    \State\quad Data $\mathcal{D}$
    \State\quad Model set $\{\mathcal{M}^k\}_{\mathcal{K}}$, partitioned into $K$ subsets $\{\mathcal{M}^k\}_{\mathcal{I}^k}$ 
    \State\quad Flux priors 
    $p(v_{\mathcal{M}_i^k} 
    \mid \mathcal{M}_i^k
    )$
    \State\quad Model set priors
    $p(\{\mathcal{M}^k\}_{\mathcal{I}^k})$ 
    \State\quad Model priors
    $p(\mathcal{M}_i^k \mid \{\mathcal{M}^k\}_{\mathcal{I}^k})$
    \State \textbf{output:}
    \State\quad Full model set-averaged flux posterior 
    $p(v_{\{\mathcal{M}^k\}_\mathcal{K}} \mid \mathcal{D})$ 
    \State\quad Model subset probabilities 
    $p( \{\mathcal{M}^k\}_{\mathcal{I}^k} \mid \mathcal{D})$
    \State\quad Single-model probabilities in view of the model subset $\{\mathcal{M}^k\}_{\mathcal{I}^k}$
    $p(\mathcal{M}^k_i \mid \{\mathcal{M}^k\}_{\mathcal{I}^k}, \mathcal{D})$
    \State\quad Single-model probabilities in view of the full model set $\{\mathcal{M}^k\}_{\mathcal{K}}$
    $p(\mathcal{M}^k_i \mid \{\mathcal{M}^k\}_{\mathcal{K}},  \mathcal{D})$
    \Procedure{\AC:}{}
        \For{$k' \gets 1$ to $K$} 
            \State $\left.\begin{aligned}[c] 
            p(v_{\{\mathcal{M}^{k'}\}_{\mathcal{I}^{k'}}} \mid \mathcal{D})
            &\quad \InlineCmt{Eq.~\eqref{eq:BMA}}
            \\
            p(\mathcal{D} \mid \{\mathcal{M}^{k'}\}_{\mathcal{I}^{k'}}) 
            &\quad \InlineCmt{Eq.~\eqref{eq:model-subset-evidences}}
            \end{aligned} \right\} \gets \text{TDNS,\,\,\SuI{}~Appendix~\ref{SI:S2_TDNS}}$ 
            \ForAll{$i' \in  \mathcal{I}^{k'}$}
                \State $p(\mathcal{M}_{i'}^{k'} \mid
                \{\mathcal{M}^{k'}\}_{\mathcal{I}^{k'}},
                \mathcal{D}) \gets 
                \frac{\# \mbox{samples} (\mathcal{M}_{i'}^{k'})}{\# \mbox{samples} (\{ \mathcal{M}^{k'}\}_{\mathcal{I}^{k'}})}$
            \EndFor
            \State \Return $p(\{\mathcal{M}^{k'}\}_{\mathcal{I}^{k'}} \mid \mathcal{D})$ 
            \Comment{Eq.~\eqref{eq:model-subset-probability}}
            \State \Return $p(\mathcal{M}_{i'}^{k'} \mid \{ \mathcal{M}^k \}_{\mathcal{K}}, \mathcal{D}) 
            \gets p(\mathcal{M}_{i'}^{k'} \mid \{\mathcal{M}^{k'}\}_{\mathcal{I}^{k'}}, \mathcal{D}) \cdot
            p(\{\mathcal{M}^{k'}\}_{\mathcal{I}^{k'} \mid \mathcal{D})}$ 
        \EndFor
    \State \Return $p(v_{\{\mathcal{M}^{k}\}_\mathcal{K}} \mid \mathcal{D})$ 
    \Comment{Eq.~\eqref{eq:BMA-with-model-sets}}
    \EndProcedure
    \end{algorithmic}
\end{algorithm}

\AC{} has the advantageous property that the loop over $k'$ (L~11) is embarrassingly parallel. This enables efficient distribution of subset-level inference across multi-core and multi-node HPC architectures. To match available compute resources the subset partitioning of the full model set can be further refined provided that models from different reaction sets remain in separate subsets. This flexibility makes \AC{} naturally scalable to large model spaces.

\subsection{Implementation}

We implemented \AC{} for \cmfa{} using four open-source packages. 
(i) For DNS we use the high-performance implementation \texttt{DNest4}~\citep{Brewer2016}. \texttt{DNest4} (v0.2.4) allows us to implement custom likelihood functions and proposal mechanisms. To improve performance for sampling \cmfa{} models, enable robust checkpointing on supercomputer systems, and support an exact representation of samples in the output (hexfloat), we created a fork of \texttt{DNest4}, available at \url{https://github.com/modsim/dnest4}. 
(ii) For convex polytope sampling we use \texttt{hopsy} (v1.7.0)~\citep{hopsy}, which builds on the \C++ library \texttt{HOPS} for optimized sampling in high-dimensional polytopes~\citep{HOPS}. We developed an adapter that connects \texttt{hopsy} and \texttt{DNest4} to reuse our RJMCMC proposals directly within \texttt{DNest4}. 
(iii) For simulation of \Ciso{} labeling data we use the high-performance simulator \texttt{13CFLUX} (v3.0.0a)~\citep{Stratmann2025}. Fast simulation is critical because likelihood evaluations for realistic \cmfa{} models dominate the computational cost of Bayesian sampling procedures. 
Although each of the above libraries provides a Python interface, we use their native \C++ APIs to minimize overhead. 
(iv) To improve sampling efficiency, we preprocess \cmfa{} models using the Python package \texttt{PolyRound} (v0.3.0)~\citep{Theorell2022}, which removes redundant constraints and computes a maximum volume ellipsoid for rounding the convex flux polytopes~\citep{Jadebeck2023}.

\section{Application studies}
\label{sec:applications}
\begin{figure*}[t]
    \centering
    \includegraphics[width=1.0\textwidth]{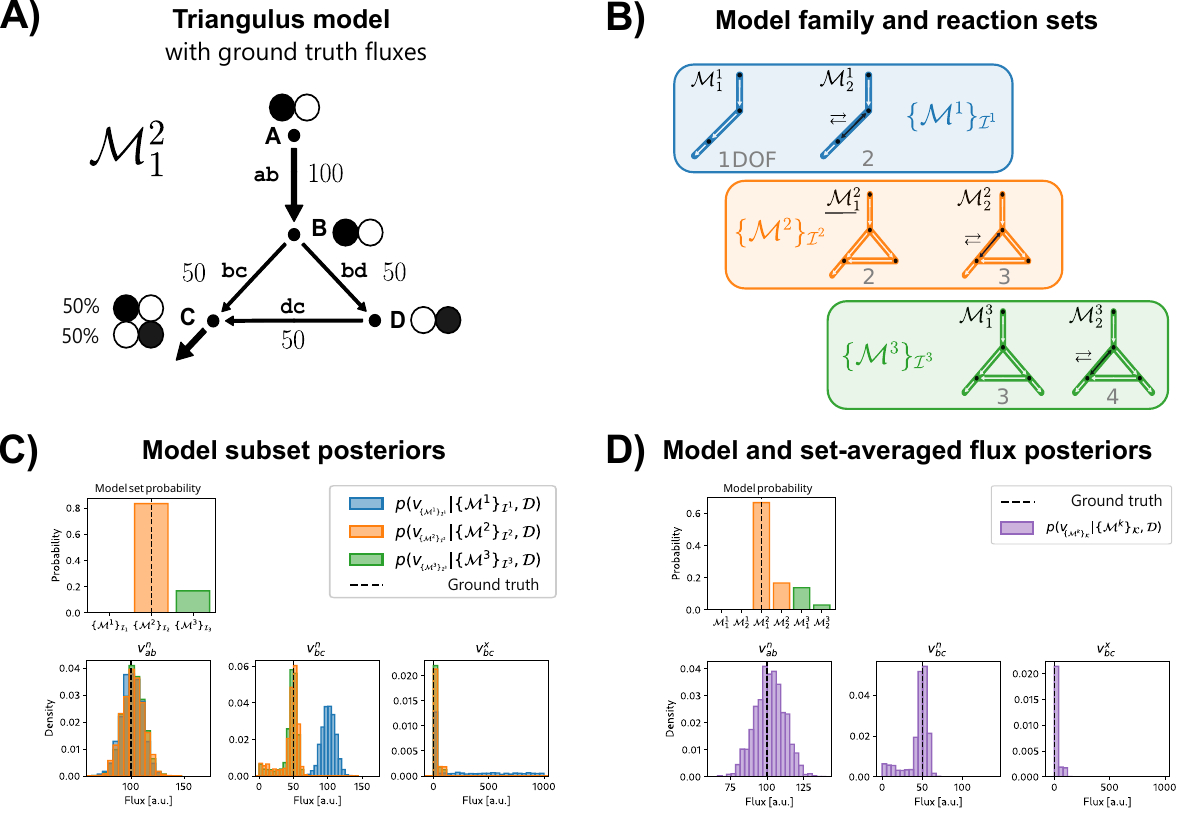}
    \caption{\textbf{\tri{} case study.}
    (A) Ground truth network with ILE data for the positionally labeled tracer (\texttt{A}\tikzcircle[black, fill=black]{2pt}\tikzcircle[black, fill=white]{2pt}). 
    Due to the scrambling reaction \texttt{bd}, the labeling pattern of \texttt{C} reflects the flux ratio between the two pathways.
    (B) Model variants and their reaction sets; within each subset models differ by uni- or bidirectionality of reaction \texttt{bc}.
    (C) Inference results at the model subset level. Subset posterior probabilities and corresponding marginal flux posteriors are color-coded. The blue subset lacks reactions \texttt{bd} and \texttt{dc}, constraining \netflux{bc} to equal \netflux{ab}.
    (D) Inference results on the full model set level, showing posterior probabilities of individual models and model-averaged marginal flux posteriors. 
    See \SuI{}~Appendix Fig.~\ref{fig:triangulus_convergence} for convergence diagnostics.} 
    \label{fig:triangulusProblemDescription}
\end{figure*}

To evaluate the \AC{} algorithm, in this work we address three questions:
\begin{enumerate}
    \item Does the method yield correct and interpretable results?
    \item Is our implementation computationally efficient for real-world problems?
    \item How does data informativeness influence model uncertainty and flux inferences?
\end{enumerate}
We investigate these aspects with two case studies: First, a minimal toy system that provides an intuitive understanding and enables validation 
under controlled conditions. 
Second, an \textit{Escherichia~coli} model~\citep{Long2019} is used to assess scalability and inference behavior under realistic levels of data informativeness.

\subsection{\normalfont Illustrative example: \itshape Triangulus}

We first apply \AC{} to a minimal network, referred to as \tri{}, capturing key features of \cmfa{} models: a parallel pathway between a source metabolite (\texttt{A}) and an observable metabolite (\texttt{C}) (cf.~\Cref{fig:triangulusProblemDescription}A). The connection proceeds either via an unidirectional two-step route via metabolite \texttt{B} (\texttt{ab}-\texttt{bc}) or an unidirectional three-step route involving  metabolites \texttt{B} and \texttt{D} (\texttt{ab}-\texttt{bd}-\texttt{dc}). While the direct route preserves atom positions, the alternative pathway introduces scrambling, making the labeling of \texttt{C} informative about the flux ratio between the net fluxes \netflux{bc} and \netflux{bd} (with \netflux{bd}=\netflux{dc}) when using a positionally labeled substrate.

Synthetic data were generated using the positionally labeled tracer \texttt{A}\tikzcircle[black, fill=black]{2pt}\tikzcircle[black, fill=white]{2pt} and a reference flux configuration with equal flux through both pathways. Mass isotopologues of the first and second \texttt{C} fragments were simulated, and domain-typical measurement noise was added to the uptake flux \netflux{ab} (10\% relative Gaussian error) and the labeling data (standard deviation of 0.01). In this setup, the data are informative for the fluxes. 

To introduce structural uncertainty, we construct a small model family (cf.~\Cref{fig:triangulusProblemDescription}B) by allowing reaction \texttt{bc} to be bidirectional and by introducing an additional export reaction for \texttt{D}. The resulting model set comprises under-complex variants ($\mathcal{M}^1_1$ DOF=1), the ground-truth \tri{} model $\mathcal{M}^2_1$ (DOF=2), a second model with two independent flux parameters ($\mathcal{M}^1_2$), and three more flexible over-parametrized model variants ($\mathcal{M}^2_2$, $\mathcal{M}^3_1$, $\mathcal{M}^3_2$ with DOF=3-4). By construction, these models form three reaction sets (blue, orange, green), which also define the \AC{} model set partitioning.
Model and flux priors for all models, as well as conventional single-model flux posteriors for reference are provided in the \SuI{} (cf.~Appendix Table~\ref{tab:priors_triangulus}, Figs.~\ref{fig:triangulus_priors} and~\ref{fig:triangulus_posteriors}).

Applying \AC, we first analyze the inferred model subset probabilities and subset-averaged flux posteriors 
(\Cref{fig:triangulusProblemDescription}C). As expected, the blue subset receives effectively zero probability, confirming that its models cannot explain the data. The orange and green subsets both provide viable explanations for the data, with the orange subset (containing the ground-truth model) dominating with $\sim$83\%, compared to $\sim$17\% for the green subset. This reflects an Ockham's razor~\citep{MacKay2008,Johnjoe2023}: although both subsets can explain the data equally, only a small fraction of the larger flux space of the more flexible models of the green subset yields flux solutions consistent with the measurements, reducing the evidence of the green subset.

Subset-averaged flux posteriors further illustrates this. The uptake flux \netflux{ab}, which is directly measured, is accurately recovered by all models, including its uncertainty. In contrast, models in the blue subset produce biased estimates for internal fluxes due to their inability to reproduce the data. The bidirectionality of reaction \texttt{bc} in $\mathcal{M}^1_2$ remains undecidable, with approximately equal posterior support for being uni- and bidirectional, reflected in the distribution of exchange flux values \xchflux{bc}.
Both the orange and green subsets explain the labeling data and therefore yield nearly identical subset-averaged flux posteriors, with most probability mass concentrated  around the ground truth values. The lower tails of the \netflux{bc} posterior arise from alternative explanations involving non-zero exchange flux values \xchflux{bc}, which introduces an alternative, though less probable, explanation of the \texttt{C} labeling measurements through label cycling.

Combining these subset-related results yields model probabilities in view of the full model set and the full model set-averaged flux posterior (cf. \Cref{fig:triangulusProblemDescription}D). The too simple models in the blue subset  receive negligible probability, while among the remaining models, the model evidence balances fit quality against model flexibility. Consequently, the ground truth model $\mathcal{M}^2_1$ dominates ($\sim$66\% probability), followed by more flexible ones, the orange model $\mathcal{M}^2_2$ ($\sim$16\%) and the two green models $\mathcal{M}^3_1$ ($\sim$14\%) and $\mathcal{M}^3_2$($\sim$3\%). 

Overall, these posterior probabilities reflect the penalization of unnecessary model flexibility, while favoring the simplest model capable of explaining the data (the ground truth model $\mathcal{M}^2_1$). As a result, the full model-averaged flux posterior closely matches the data-generating flux configuration. This example illustrates that, for sufficiently informative data, \AC{} yields interpretable results, balances model fit and complexity, and recovers the ground-truth.  

\subsection{\normalfont Realistic example: \itshape Escherichia~coli}

\begin{figure*}[ht]
    \centering
    \includegraphics[width=1.0\textwidth]{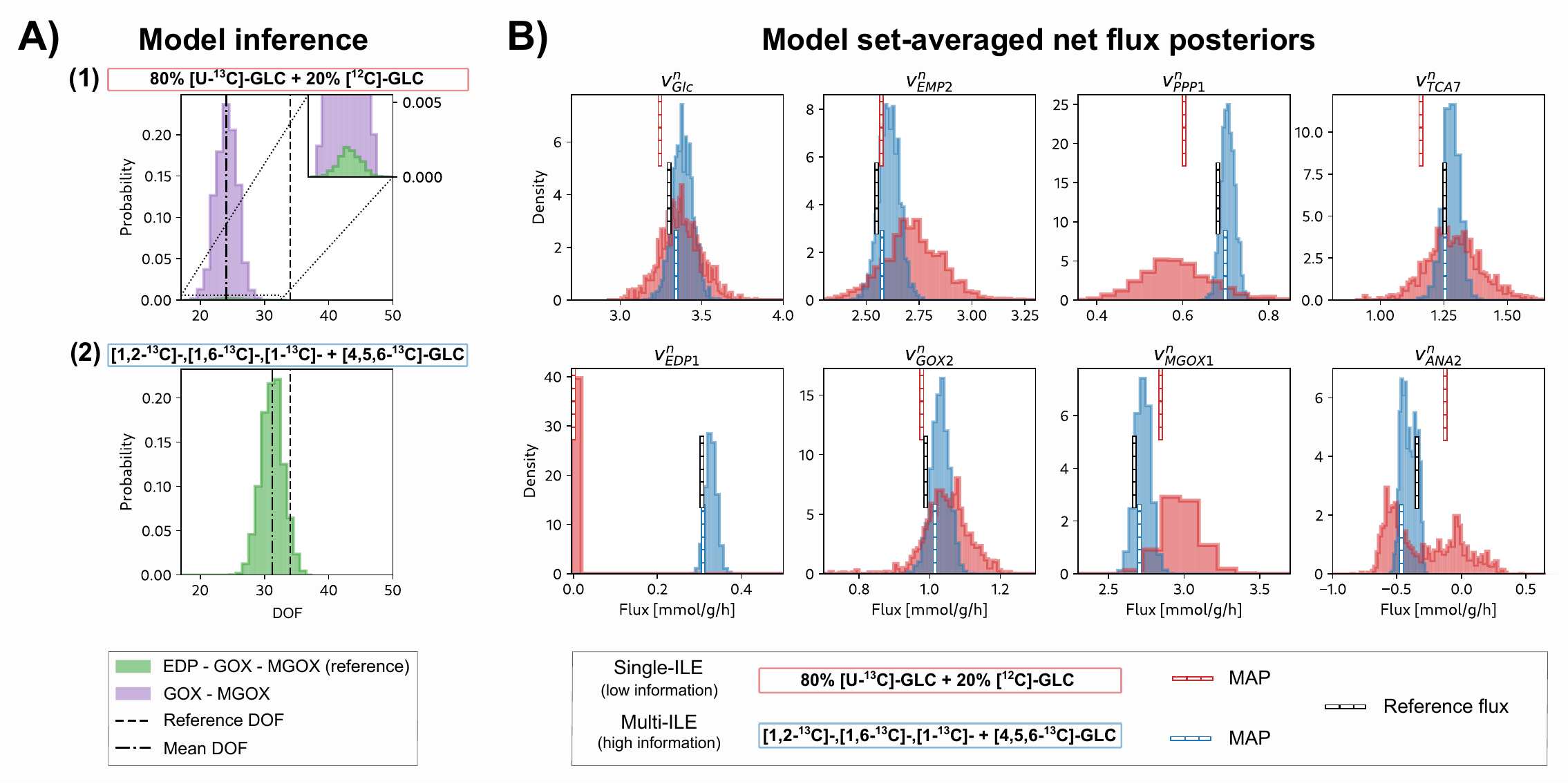}
    \caption{
        \textbf{Inference results for the \ECfull{} case study.}
        (A) Model inference results shown as a stacked histogram over effective model DOFs. Each bar represents the total posterior probability of all model variants with a given number of DOF, partitioned by model subset (color-coded). 
        For the single-ILE evaluation, only two subsets receive non-negligible probability: the subset containing the reference model is strongly down-weighted (\SI{\sim0.9}\percent), while the simpler model subset with GOX and MGOX included, and TPI and EDP excluded dominates (\SI{99.1}\percent).
        The DOF of the reference model is indicated by the dashed line. 
        For the multi-ILE analysis (lower panel), the reference reaction set is recovered with $>$ \SI{99.9}\percent probability.
        (B) Marginal model set-averaged posterior distributions for selected net fluxes, including the maximum a posteriori (MAP) estimates. Results for all 75 net fluxes are provided in \SuI{}~Appendix Fig.\ref{fig:overview}. Inferences are reproducible across \AC{} runs (cf.~\SuI{}~Appendix~\ref{ssec:reproducibility}).
    }
    \label{fig:crown_net_fluxes}
\end{figure*}

Although the \tri{} example provides useful intuition, it does not reflect the complexity of practical \cmfa{} problems, where models comprise dozens to hundreds of reactions, combinatorial model spaces may reach millions or more variants, and experimental data are/is often insufficient to resolve all fluxes and inform a unique model structure. To assess the performance of \AC{} under such conditions, we apply it to a published \cmfa{} study by~\citet{Long2019}. 
The authors analyzed an \ECfull{} $\Delta$\textit{tpiA} mutant, in which deletion of the triosephosphate isomerase gene results in nearly zero residual \texttt{TPI} activity. To avoid dihydroxyacetone phosphate accumulation, the strain activates the methylglyoxal (MGOX) pathway. While the glyoxylate shunt (GOX) is clearly active, the Entner-Doudoroff pathway (EDP) carries only a small net flux. The system therefore constitutes a challenging test case with multiple latent pathways in central metabolism. 

We reconstructed the published metabolic network, but with both \netflux{\textit{TPI}} and \xchflux{\textit{TPI}} (denoted TPI pathway) fixed to zero (\SuI{}~Appendix~Fig.~\ref{fig:model}). This reference model comprises \num{81} metabolites and \num{126} reactions, with \num{11} independent net fluxes and \num{23} independent exchange fluxes (DOF=\num{34}). Model uncertainty was introduced by treating the activity of the TPI, EDP, GOX, and MGOX pathways as uncertain. This yields twelve reaction sets: when TPI is active, MGOX, EDP and GOX may be active or inactive (eight reaction sets), whereas when TPI is inactive, MGOX must remain active while EDP and GOX can be on or off (four reaction sets); see also \SuI{}~Appendix~Table~\ref{tab:modelsubsets_ecoli}. Combined  with uncertain reaction bidirectionality, this gives \num{46976204800} alternative model structures, spanning a DOF-range from \num{9} to \num{46}. 
This defines an application-scale benchmark: data are generated from a ground truth, while inference is performed over a combinatorial model family reflecting realistic pathway-level uncertainty. As before, we partition the model space into subsets corresponding to reaction sets (cf.~\SuI{}~Appendix~Table~\ref{tab:modelsubsets_ecoli}) and specify priors, reported in \SuI{}~Appendix~Sec.~\ref{ssec:priorspec}.

\subsubsection{Single-ILE evaluation}

We first consider a typical experimental scenario using a single, relatively uninformative tracer mixture (\tracerA{} glucose). Synthetic data were generated from the reference model and perturbed using Gaussian noise as reported in the original study.

\Cref{fig:crown_net_fluxes}A.1 summarizes the model inference in terms of DOF distribution over the models that carry the majority of posterior probability mass. Specifically, we introduce the \textit{effective model set} (EMS) as the smallest subset of models whose cumulative posterior probability exceeds \num{95}\%, thereby effectively contributing to the model set-averaged flux posterior. Despite the large candidate space, the EMS contains only \num{1572} models (\SI{3.3e-6}{\percent}), indicating strong concentration of posterior mass. All effective models share a core reaction set with active GOX and MGOX pathways (plum), which receives \SI{99.1}{\percent} posterior probability, whereas the reference reaction set including active EDP, MGOX, and GOX pathways receives only \SI{\sim0.9}{\percent}. Thus, for the applied tracer, EDP activity is not required to explain the data. 

This outcome directly reflects limited data informativeness: because the model evidence averages the likelihood over the prior, additional DOFs are penalized unless supported by the data (Ockham's razor). Here, the labeling data do not provide sufficient information to  support the additional flexibility associated with the EDP, leading to a down-weighting of models in which the EDP is active. Importantly, this reflects a lack of statistical support under the given data and priors, rather than evidence of biological inactivity.
Consistent with this interpretation, the mean DOF of the EMS is substantially reduced compared to the reference model (\SI{24.1(1.6)} vs. \num{34}). Obviously, the deactivation of the EDP pathway accounts for only one of the reduced DOFs; the remaining \num{\sim 9} DOFs are attributed to unresolved bidirectionalities. On average, only  \SI{14.1(1.6)} of the \num{23} bidirectional reactions in the reference model are inferred as bidirectional. This quantifies the limited information content of the data and illustrates how BMA deals with structural model uncertainty rather than neglecting it.

The corresponding model set-averaged marginal flux posteriors in \Cref{fig:crown_net_fluxes}B show mostly unimodal distributions for representative core fluxes (cf.~\SuI{}~Appendix~Fig.~\ref{fig:overview} for the remaining net fluxes), with exception of the malate dehydrogenase (\texttt{ANA2}) flux, where a bimodal shape indicates alternative anaplerotic modes, a phenomenon previously observed theoretically~\citep{Kappelmann2016} and in practice~\citep{BorahSlater2023}. 
Due to the dominance of models without EDP in the EMS, central fluxes in the EMP and PPP are shifted relative to the reference fluxes. Nevertheless, uncertainty bounds generally encompass the reference fluxes. Interestingly, compared to single-model inference, uncertainties under full model uncertainty for several core fluxes in EMP and PPP are reduced (cf.~\SuI{}~Appendix~Fig.~\ref{fig:singleA}). This arises because the data favor models with fewer DOFs, effectively restricting the admissible flux space.

The \AC{} run converged after \num{147456} core-hours (\num{4} days) on the JURECA supercomputer (cf.~\SuI{}~Appendix~Fig.~11), demonstrating computationally feasibility at application-scale. 
In summary, under a single, weakly informative ILE dataset, \AC{} favors simpler models (Ockham's razor), down-weights unsupported pathways, and propagates the uncertainty from the model structures into flux inferences without inflating uncertainty beyond what is consistent with the data and the priors. 

\subsubsection{Multi-ILE evaluation}

We next investigate how increased data informativeness affects the inferences. To this end, we repeated the analysis using three ILE datasets with complementary tracers (\tracerBa{}, \tracerBb{}, and \tracerBc{}). Each tracer resolves different parts of central metabolism, and their joint analysis triples the number of measurements while leaving the models' DOF unchanged. We therefore expect closer reflection the reference model and more precise flux estimates. 

When analyzing the three datasets jointly, the inference landscape changes markedly (cf.~\Cref{fig:crown_net_fluxes}A.2 and B). The EMS contains \num{1205} models (\SI{2.6e-6}{\percent} of all variants), a reduction of \SI{\sim 23}{\percent} compared to single-ILE case. Most notably, the reference reaction set (EDP-GOX-MGOX) is now recovered with overwhelming probability ($>$\SI{99.9}{\percent}). 
Consistently, the DOF distribution shifts toward higher values 
and centers at \SI{31.2(1.7)}, approaching the reference DOF (\num{34}). This confirms that the additional data provide sufficient evidence to support the previously unresolved EDP pathway activity and reaction bidirectionality. Interestingly, we found a slightly increased DOF spread compared to the single-ILE case which reflects residual uncertainty in the two unresolved reaction bidirectionalities.

Flux posteriors reflect the increase in data informativeness. Compared to the single-ILE case, they are substantially sharper and align closely with the reference flux map (cf.~\SuI{}~Appendix~Fig.~\ref{fig:overview}). In particular, the previously bimodal \texttt{ANA2} net flux resolves into a single mode, indicating that the additional data eliminate competing metabolic explanations. 
Together, these results highlight the central role of data informativeness: while the single dataset supports only simplified models, the multi-dataset BMA analysis recovers the reference pathway structure and reduces both structural and flux uncertainty. 
Notably, this sharpening occurs despite explicitly accounting for model uncertainty. The multi-ILE analysis was obtained using \num{405504} core-hours (\num{11} days) on JURECA (cf.~\SuI{}~Appendix~Fig.~11), demonstrating computational scalability to integrate multiple data sets. 

Overall these experiments show how BMA as implemented in \AC{} adapts the effective model set complexity to the information content of the data: unsupported flexibility is penalized, while informative data enable recovery of metabolic pathways and precise, yet not over-confident flux estimates (cf.~\SuI{}~Appendix~Fig.~\ref{fig:singleA} for a comparison with conventional single-model inference). 
Posterior model probabilities thus provide a principled measure of relative support within the specified model family, enabling the identification of pathways that are statistically supported by the data.

\section{Conclusion}
\label{sec:conclusion}
We presented \AC{}, a scalable BMA algorithm for \cmfa{} that quantifies metabolic fluxes while explicitly accounting for pathway- and reaction-level model uncertainty through evidence-based model-set weighting. Rather than relying on ad hoc model selection, \AC{}  integrates over a spectrum of plausible models, yielding interpretable posterior support for pathway hypotheses and flux posteriors that reflect both parameter and structural uncertainty. 

Importantly, \AC{} does not eliminate uncertainty nor guarantees identification of a single "correct" model; instead, it manages structural uncertainty in a statistically principled manner. By propagating both data and model uncertainty into the flux inferences, it makes explicit which aspects of the model are supported by the data and which remain unresolved. 
Algorithmically, \AC{} combines RJMCMC for trans-dimensional model exploration with DNS for robust evidence estimation, enabling principled inferences in large, nonlinear, and complex-constrained settings. This is particularly relevant for \cmfa{}, where uncertain pathways and reaction bidirectionalities can otherwise lead to model-dependent conclusions. Consistent with evidence-based inference,  model parts with weak data support may be down-weighted despite being biologically active, reflecting limited identifiability in view of the data and prior rather than biological absence. Conversely, increasing data informativeness leads to recovery of the correct model configuration and sharper flux without over-committing to unsupportable assumptions.

We demonstrated correctness and interpretability on a toy example and computational feasibility on an application-scale \EC{} benchmark spanning billions of model variants. Owing to its divide-and-conquer formulation, our \AC{} implementation is naturally parallelizable, and its computational cost can be adapted through model set partitioning and algorithmic tuning. While computationally demanding, it provides information inaccessible to single-model analyses, a trade-off that is particularly justified in \cmfa{}, where experimental data acquisition is costly and limited.

Together, these results establish \AC{} as a practical and extensible framework for model misspecification-aware flux inference. By explicitly accounting for structural uncertainty, it enhances the transparency, robustness, and interpretability of \cmfa{} analyses and provides a principled basis for uncertainty-aware quantitative metabolic studies.

\section{Competing interests}
\noindent
No competing interest is to be declared.

\section{Acknowledgments}
The authors are thankful to Axel Theorell for the Bayesian spirit, Wolfgang Wiechert for excellent working conditions at the IBG-1 and acknowledge the computing time on the supercomputer JURECA at Forschungszentrum Jülich (grant no. \texttt{hpcmfa}).

\section{Funding}
This work was performed as part of the Helmholtz School for Data Science  in  Life,  Earth,  and  Energy  (HDS-LEE)  and  received  funding  from  the  Helmholtz  Association  of  German Research Centres.

\bibliographystyle{abbrvnat}


\newpage
\appendix
\onecolumn

{
    \centering
    \textbf{Supplementary Information for:}\\[3ex]
    \textbf{\Large 
        Trans-dimensional Bayesian model averaging for \textsuperscript{13}C-based metabolic flux analysis: Evidence-based flux inference under structural model uncertainty
        \vspace*{\baselineskip}\\[5ex]
    }
}


%
%
\section{\, BMA formulation for model sets}
\label{SI:S1_BMA_modelsets}

The BMA formulation to determine the model set-averaged flux posterior $v_{\{\mathcal{M}^k\}_\mathcal{K}}$ in view of the full model set $\{\mathcal{M}^k\}_\mathcal{K}$
\begin{equation}
   p( v_{\{\mathcal{M}^k\}_\mathcal{K}} \mid  
   \mathcal{D}) 
    = 
    \sum\limits_{k'=1}^K
    p( \{\mathcal{M}^{k'}\}_{\mathcal{I}^{k'}} \mid 
    \mathcal{D}) 
    \cdot 
    p(v_{\{\mathcal{M}^{k'}\}_{\mathcal{I}^{k'}}} \mid 
     \mathcal{D})
    \label{eq:BMA-with-model-sets_SI}
\end{equation}
is equivalent to
\begin{equation} 
    p( v_{\{ \mathcal{M}^k \}_\mathcal{K}} \mid \mathcal{D}) = 
    \sum\limits_{\substack{k' = 1,\\i' \in \mathcal{I}^{k'}}}^K
        p( \mathcal{M}^{k'}_{i'} \mid  \mathcal{D} ) 
        \cdot  p( v_{\mathcal{M}^{k'}_{i'}} \mid \mathcal{D})
    \label{eq:full-BMA-double-sum} 
\end{equation}
where in contrast to the main text, we have suppressed the conditioning of $p(\mathcal{M}_{i'}^{k'}|\mathcal{D})$ on the (fixed) model set $\{\mathcal{M}^k\}_{\mathcal{K}}$.

\noindent To prove the equivalence of Eq.~\eqref{eq:BMA-with-model-sets_SI} and Eq.~\eqref{eq:full-BMA-double-sum}, we condition the probability of all models $\mathcal{M}_{i}^{k}$ on their containing model subset $\{\mathcal{M}^k\}_{\mathcal{I}^k}$ 
\begin{equation} 
    p(\mathcal{M}^{k}_{i}\mid \mathcal{D}) 
    = 
    p(\{\mathcal{M}^{k}\}_{\mathcal{I}^{k}}\mid \mathcal{D})
    \cdot 
    p(\mathcal{M}^{k}_{i} \mid \{\mathcal{M}^{k}\}_{I^{k}},\mathcal{D}),\ \forall i\in\mathcal{I}^k, k\in\{1,\dots,K\}
    \label{eq:conditional} 
\end{equation}

\noindent It then follows
\begin{eqnarray*}
    p( v_{\{\mathcal{M}^k\}_\mathcal{K}} \mid \mathcal{D}) 
    & \overset{Eq.~\eqref{eq:full-BMA-double-sum}}{=} & 
    \sum \limits_{\substack{k' = 1 \\ i' \in \mathcal{I}{k'}}}^K
        p( \mathcal{M}^{k'}_{i'} \mid \mathcal{D} ) 
        \cdot  p( v_{\mathcal{M}^{k'}_{i'}} \mid \mathcal{D}) \\
    & \overset{Eq.~\eqref{eq:conditional}}{=} &    
    \sum\limits_{\substack{k' = 1 \\ i' \in \mathcal{I}^{k'}}}^K
        p( \{ \mathcal{M}^{k'} \}_{\mathcal{I}^{k'}} \mid \mathcal{D}) 
        \cdot
        p( \mathcal{M}^{k'}_{i'} \mid \{ \mathcal{M}^{k'} \}_{\mathcal{I}^{k'}}, \mathcal{D} ) 
        \cdot
        p( v_{\mathcal{M}^{k'}_{i'}} \mid \mathcal{D}) \\
    & = &
    \sum\limits_{k' =1 } ^ K
        p( \{ \mathcal{M}^{k'} \}_{\mathcal{I}^{k'}} \mid \mathcal{D} ) \cdot
        \sum\limits_{i' \in \mathcal{I}^{k'}} 
            p( \mathcal{M}^{k'}_{i'} \mid \{ \mathcal{M}^{k'} \}_{\mathcal{I}^{k'}}, \mathcal{D} ) 
            \cdot
            p( v_{\mathcal{M}^{k'}_{i'}} \mid \mathcal{D}) \\
    & \overset{Eq.~\eqref{eq:BMA}}{=} &
    \sum\limits_{k' = 1}^K
        p( \{ \mathcal{M}^{k'} \}_{\mathcal{I}^{k'}} \mid \mathcal{D})
        \cdot
        p( v_{\{ \mathcal{M}^{k'} \} _{\mathcal{I}^{k'}}} \mid \mathcal{D})
\end{eqnarray*}
where we use the BMA formulation for the model set $\{\mathcal{M}\}_{\mathcal{I}^k}$ in Eq.~(1) of the main manuscript, yielding the equivalence of Eq.~\eqref{eq:BMA-with-model-sets_SI} and Eq.~\eqref{eq:full-BMA-double-sum}.

\FloatBarrier
\clearpage
%
%
\section{\, Trans-dimensional diffusive nested sampling}
\label{SI:S2_TDNS}

The computational core of \ACfull{} (\AC) given in Algorithm 1 in the main text builds on trans-dimensional diffusive nested sampling (TDNS). We adapt the TDNS algorithm in~\cite{realTDNS} to the mixed discrete-continuous flux inference problem over model subsets $\{\mathcal{M}^{k}\}_{\mathcal{I}^{k}}$. While Algorithm 1 in the main text describes how the results for the model subsets are combined to the model set-averaged flux posterior over the full model set, we here concentrate on the results per model subset. For brevity, we speak of the model set $\{\mathcal{M}^{k}\}_{\mathcal{I}^{k}}$ instead of model subset in the following.

\noindent 
Given observed data $\mathcal{D}$ and the model set $\{\mathcal{M}^{k}\}_{\mathcal{I}^{k}}$, TDNS for \AC{} provides 
\begin{itemize}
    \item 
    \textit{model set-averaged flux posterior}
        $p( v_{\{\mathcal{M}^{k}\}_{\mathcal{I}^{k}}} \mid \mathcal{D})$
    \item \textit{model set evidences}
        $p(\mathcal{D} \mid \{\mathcal{M}^{k}\}_{\mathcal{I}^{k}})$ 
\end{itemize}
These quantities enable the computation of posterior model set probabilities and the full Bayesian model-averaged flux posteriors as described in the main text. 
As by-product, samples approximating the single-model posterior $p(\mathcal{M}_i^k \mid \{\mathcal{M}^k\}_{\mathcal{I}^k},\mathcal{D})$ in view of the considered model set $\{\mathcal{M}^k\}_{\mathcal{I}^k}$ are generated.
\\

\noindent
The TDNS for \AC{} algorithm combines ideas from diffusive nested sampling (DNS) and reversible-jump Markov chain Monte Carlo (RJMCMC), exploiting the specific structure of the flux inference problem. 
In next sections, we summarize the essential underlying concepts:
(i) the state space for RJMCMC, which we call the augmented flux space (\Cref{sec:state-space}), and
(ii) the model set evidence integral over the augmented flux space (\Cref{sec:model-set-evidence}).
In \Cref{sec:nested-sampling-evidence}, we introduce necessary background to nested sampling, in particular, the integral transform required to approximate the integral of the model set evidence. Finally, we describe TDNS for \AC{} in in \Cref{sec:TDNS}, along with pseudocode. 

\subsection{The augmented flux space for reversible-jump MCMC}
\label{sec:state-space}

The model set $\{\mathcal{M}^k\}_{\mathcal{I}^k}$ consists of models $\mathcal{M}_i^k$ with the same reactions, but varying reaction bidirectionality~\cite{Theorell2020}. Therefore, each model is equipped with a model-specific set of (independent) fluxes $v_{\mathcal{M}_i^k}$ that are defined on the model-specific flux polytope $\mathcal{P}_{\mathcal{M}_i^k}$. The dimension of the flux polytope, which corresponds to the DOF, may vary between different models, depending on their bidirectionality setting. 
By zeroing exchange fluxes for unidirectional reactions, the fluxes $v_{\mathcal{M}_i^k}$ can be embedded in a common ambient flux representation, while the DOF remains specific to the model.

\noindent 
TDNS uses RJMCMC~\cite{Theorell2020} to traverse the discrete-continuous space of models and model-specific fluxes. Before we describe TDNS, we formally introduce the discrete-continuous space of models and model-specific fluxes. We define the \textit{augmented flux space} $\Omega_{\{\mathcal{M}^k\}_{\mathcal{I}^k}}$ as disjoint union over the Cartesian products of all single models and their associated flux polytopes
\begin{equation}
    \Omega_{\{\mathcal{M}^k\}_{\mathcal{I}^k}} =
        \bigcup_{i\in \mathcal{I}^k } \{i\} \times \mathcal{P}_{\mathcal{M}_i^k}
    \label{eq:Omega}
\end{equation}
An element of the augmented flux space is thus represented as a tuple $(i, v_{\mathcal{M}_i^k})$ with $i \in \mathcal{I}^k$ the model index and fluxes $v_{\mathcal{M}_i^k}$ in the associated flux polytope. \\

\subsection{Model set evidence}
\label{sec:model-set-evidence}

The evidence of the model set $\{\mathcal{M}^k\}_{\mathcal{I}^k}$ is given by the integral of the model set specific likelihood $\mathcal{L}_{\{{\mathcal{M}^k}\}_{\mathcal{I}^k}}$ with respect to the prior measure $\mu_{\{\mathcal{M}^k\}_{\mathcal{I}^k}}$ related to the model set.
For this, we formally extend the single model likelihoods defined on the model-specific polytopes, to the model set related likelihood $\mathcal{L}_{\{{\mathcal{M}^k}\}_{\mathcal{I}^k}}$ on the augmented flux space $\Omega_{\{\mathcal{M}^k\}_{\mathcal{I}^k}}$ in the sense that for each element of $\Omega_{\{\mathcal{M}^k\}_{\mathcal{I}^k}}$ there is a well-defined single model likelihood. \\

\noindent
The model set-related prior measure $\mu_{\{\mathcal{M}^k\}_{\mathcal{I}^k}}$ is then defined by accumulating the priors of the single models that constitute the model set, weighted by the models' priors given the model set
\begin{equation}
\begin{split}
    &\mu_{\{\mathcal{M}^k\}_{\mathcal{I}^k}} : \Omega_{\{\mathcal{M}^k\}_{\mathcal{I}^k}} \to [0,1]
    \text{ with } \\
    &\mu_{\{\mathcal{M}^k\}_{\mathcal{I}^k}}(A) = 
    \sum \limits_{i\in\mathcal{I}^k} p(\mathcal{M}_i^k\mid \{\mathcal{M}^k\}_{\mathcal{I}^k}) 
    \cdot 
    \int_{\mathcal{P}_{\mathcal{M}^k_i}}
    \mathbb{I}_A(i, v_{\mathcal{M}_i^k})
    \cdot p(v_{\mathcal{M}_i^k}\mid\mathcal{M}_i^k) \, dv_{\mathcal{M}_i^k}
\end{split}
\label{eq:1}
\end{equation}
with $A$ a measurable subset of $\Omega_{\{\mathcal{M}^k\}_{\mathcal{I}^k}}$ and the indicator function $\mathbb{I}_A$ being $1$ for all tuples $(i, v_{\mathcal{M}_i^k}) \in A$ and $0$ else.

\noindent
Given the prior measure, the evidence of a model set $\{\mathcal{M}^k\}_{\mathcal{I}^k}$ is given by
\begin{equation}
    p(\mathcal{D}|\{\mathcal{M}^k\}_{\mathcal{I}^k}) =
   \int_{
     \Omega_{\{\mathcal{M}^k\}_{\mathcal{I}^k}}
   } 
   \mathcal{L}_{\{{\mathcal{M}^k}\}_{\mathcal{I}^k}}
   (i, v_{\mathcal{M}^k_i}) 
   \,
   d\mu_{\{\mathcal{M}^k\}_{\mathcal{I}^k}}
   (i,v_{\mathcal{M}^k_i})
   \label{eq:model-set-evidence}
\end{equation}
where $\mathcal{L}_{\{{\mathcal{M}^k}\}_{\mathcal{I}^k}}(
i, v_{\mathcal{M}_i^k})$ is the likelihood of model $\mathcal{M}_i^k$ evaluated for fluxes $v_{\mathcal{M}_i^k}$.

\subsection{Nested sampling background}
\label{sec:nested-sampling-evidence}
\begin{figure}[ht!]
    \centering
    \includegraphics[width=1\linewidth]{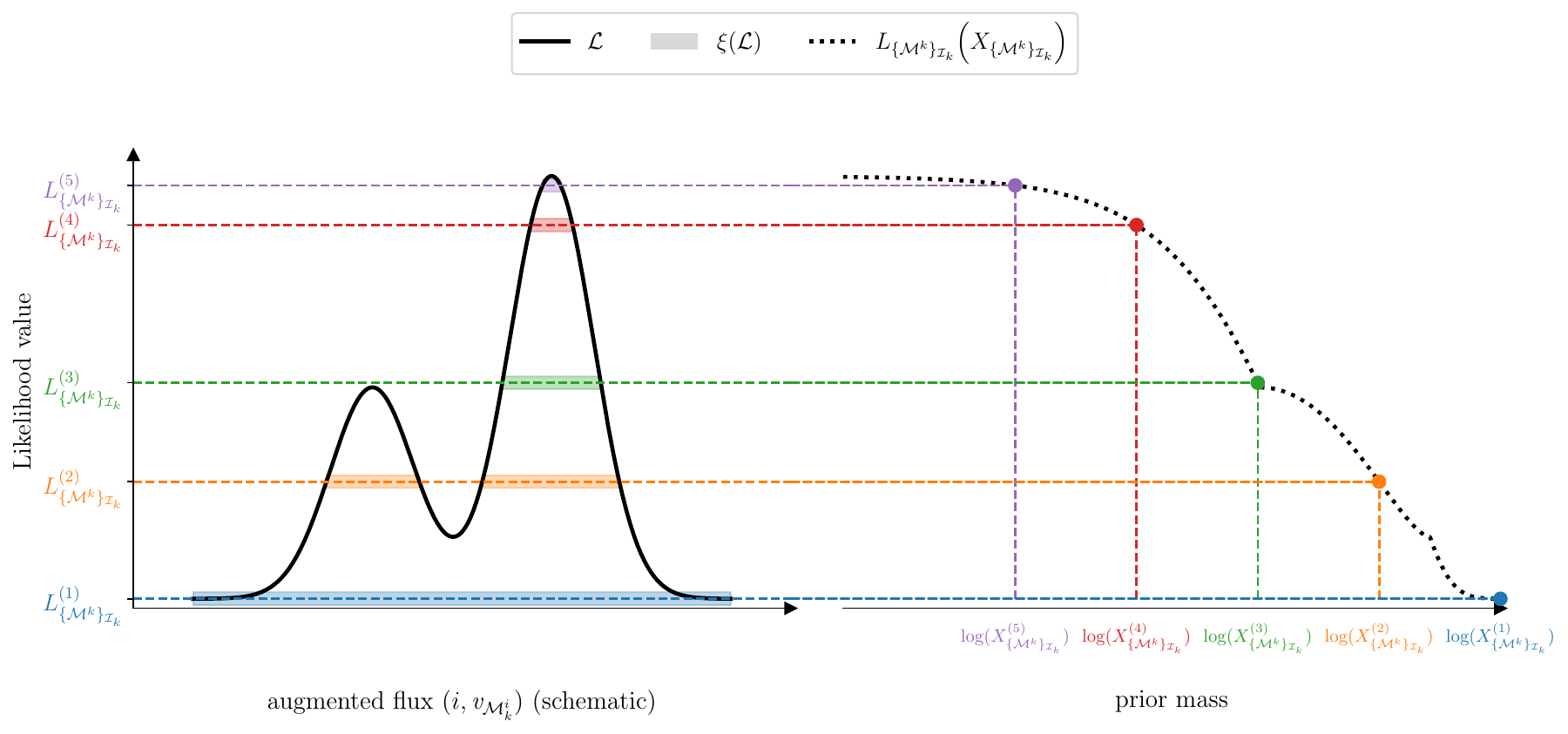}
    \caption{
    \textbf{Schematic representation of the evidence integral transform that nested sampling exploits.}
    Left: Likelihood (black solid) shown as a function of model-specific fluxes. The fluxes are here formally represented as a tuple, being an element of the augmented flux space.
    Colored domains represent the surviving prior mass $\xi$ for different likelihood values $L_{\{\mathcal{M}^k\}_{\mathcal{I}^k}}$. The length of the colored domains corresponds to their prior measure $X_{\{\mathcal{M}^k\}_{\mathcal{I}^k}}$. 
    In nested sampling, it is common practice to show the likelihood in log space. For this representation, we have chosen to display the likelihood value rather than the log likelihood. \\
    Right: Likelihood (black dotted) as a function of the prior mass. The likelihood levels on the left correspond to different values of the prior-mass coordinate $X_{\{\mathcal{M}^k\}_{\mathcal{I}^k}}$ (colored dots, prior-mass coordinates are depicted in log space).
    Kinks in the transformed likelihood emerge at likelihood values, where the likelihood goes from unimodal
    to bimodal and vice versa. These kinks are known as ``phase-transitions'' and occur when the surviving prior mass $\xi$ changes sharply at a certain likelihood level.
    }
    \label{fig:TDNS} 
\end{figure}

Determining the model set evidence in Eq.~\eqref{eq:model-set-evidence} is computationally challenging. To make the calculation tractable, the key idea of nested sampling is to transform the high-dimensional evidence integral into a one-dimensional integral~\cite{Skilling2006}. Figure~\ref{fig:TDNS} provides a conceptual visualization of the underlying transformation.\\

\noindent
Following the standard nested sampling formulation in \citep{Ashton2022}, we rewrite the model set evidence in Eq.~\eqref{eq:model-set-evidence} as 
\begin{equation}
    p(\mathcal{D}|\{\mathcal{M}^k\}_{\mathcal{I}^k}) =
    \int_0^1 
    {L}_{\{\mathcal{M}^k\}_{\mathcal{I}^k}}
    (X_{\{\mathcal{M}^k\}_{\mathcal{I}^k}} ) \, 
    dX_{\{\mathcal{M}^k\}_{\mathcal{I}^k}}
    \label{eq:1-d}
\end{equation}
with the so-called prior mass variable $X_{\{\mathcal{M}^k\}_{\mathcal{I}^k}} \in [0,1]$ and 
${L}_{\{\mathcal{M}^k\}_{\mathcal{I}^k}}$ the model set related likelihood expressed as a function of the prior mass variable
\begin{equation}
    \begin{split}
        &{L}_{\{\mathcal{M}^k\}_{\mathcal{I}^k}}:[0,1]\to[0, \infty) \quad \text{with} \\
        &{L}_{\{\mathcal{M}^k\}_{\mathcal{I}^k}}
        ( X_{\{\mathcal{M}^k\}_{\mathcal{I}^k}})
        =
        \sup\{\lambda \in 
        \text{im}(
            \mathcal{L}_{\{{\mathcal{M}^k}\}_{\mathcal{I}^k}}
            )
            \mid
            \xi(\lambda) > X_{\{\mathcal{M}^k\}_{\mathcal{I}^k}}\} 
        \end{split}
\end{equation}
where $\text{im}(\mathcal{L}_{\{{\mathcal{M}^k}\}_{\mathcal{I}^k}})$ consists of all likelihood values for any element of the augmented flux space, and $\xi$ the so-called surviving prior-mass function
\begin{equation}
    \begin{split}
       &\xi: [0, \infty) \to [0,1] \text{ with} \\
       &\xi(\lambda)=
           \mu_{\{\mathcal{M}^k\}_{\mathcal{I}^k}}
        (\{
        (i,v_{\mathcal{M}^k_{i}}) \in
        \Omega_{\{\mathcal{M}^k\}_{\mathcal{I}^k}}
        \mid 
            \mathcal{L}_{\{{\mathcal{M}^k}\}_{\mathcal{I}^k}} (i, v_{\mathcal{M}^k_{i}}) > \lambda
        \})
    \end{split}
\end{equation}

\noindent
Eq.~\eqref{eq:1-d} holds in the general case of measurable bounded likelihoods~\cite{schittenhelm2021nestedsamplinglikelihoodplateaus}, which is fulfilled for the case of \cmfa.


\subsection{TDNS for \AC{}}
\label{sec:TDNS}

This section describes TDNS for \AC{} for Bayesian model averaging in \cmfa. We refer to the original literature on (trans-dimensional) DNS for technical details ~\cite{Brewer2011-kq, brewer2015inferencetransdimensionalbayesianmodels}. Here, we focus on how the two core results, model set-averaged flux posterior samples and model set evidence, are computed
\begin{enumerate}
    \item an approximation to the $1$-dimensional evidence integral Eq.~\eqref{eq:1-d}, and 
    \item model set-averaged flux samples $v^{(t)}_{\{\mathcal{M}^k\}_{\mathcal{I}^k}}$ to approximate the posterior $p(v_{\{\mathcal{M}^k\}_{\mathcal{I}^k}} \mid \mathcal{D}) $ 
\end{enumerate}
The following section introduces the notation and definitions required to understand TDNS for \AC{}, as described in the algorithm listing ~\ref{alg:TDNS}. Finally, we briefly discuss parameter choices, and how to diagnose TDNS for \AC{} runs.\\

\subsubsection{Solving the evidence integral}

To approximately solve Eq.~\eqref{eq:1-d}, TDNS constructs a ladder of $J_{\max}\in\mathbb{N}$ likelihood levels 
\begin{equation}
    0=L_{\{\mathcal{M}^k\}_{\mathcal{I}^k}}^{(1)} < \cdots < 
    L_{\{\mathcal{M}^k\}_{\mathcal{I}^k}}^{(j)} < \cdots < 
    L_{\{\mathcal{M}^k\}_{\mathcal{I}^k}}^{(J_{\max})}    
    \label{eq:ll}
\end{equation}
and estimates the prior masses 
\begin{equation}
    1=X_{{\{\mathcal{M}^k}\}_{\mathcal{I}^k}}^{(1)} > \cdots > 
    X_{{\{\mathcal{M}^k}\}_{\mathcal{I}^k}}^{(j)} > \cdots > X_{{\{\mathcal{M}^k\}_{\mathcal{I}^k}}}^{(J_{\max})} 
\end{equation}
such that
\begin{equation}
    {L}_{\{\mathcal{M}^k\}_{\mathcal{I}^k}}(
    X_{{\{\mathcal{M}^k}\}_{\mathcal{I}^k}}^{(j)}
    )
    = {L}_{\{\mathcal{M}^k\}_{\mathcal{I}^k}}^{(j)}, \, \forall j
\end{equation}

\noindent
For each single model $\mathcal{M}_i^k$, the joint prior over its fluxes in view of the considered model set $\{\mathcal{M}^k\}_{\mathcal{I}^k}$ is given by 
\begin{equation}
    p(i,v_{\mathcal{M}_i^k}) = 
    p(v_{\mathcal{M}_i^k} \mid \mathcal{M}_i^k) \cdot 
    p(\mathcal{M}_i^k \mid \{\mathcal{M}^k\}_{\mathcal{I}^k})   
\end{equation}
With this, we formulate the likelihood constrained prior probability for each model-flux combination given the $j$-th likelihood level as the prior-mass-weighted joint probability
\begin{equation}
    p(i, v_{\mathcal{M}_i^k} \mid j) = 
    \begin{cases}
        \frac{p(i, v_{\mathcal{M}_i^k})}{X^{(j)}_{{\{\mathcal{M}^k\}_{\mathcal{I}^k}}}} , & \text{if} \,\,
       \mathcal{L}_{\{{\mathcal{M}^k}\}_{\mathcal{I}^k}}(
       i, v_{\mathcal{M}_i^k})>
        {L}^{(j)}_{\{\mathcal{M}^k\}_{\mathcal{I}^k}} \\
        0,& \text{otherwise}
    \label{eq:likelihood-constrained-priors}
    \end{cases}
\end{equation}

\noindent
The likelihood levels in Eq.~\eqref{eq:ll} are constructed iteratively. Let $J$ be the number of likelihood levels created so far. Following~\cite{Brewer2011-kq}, we choose the probability of level $j$ given the number of so far created levels $J$ as follows
\begin{equation}
    p_J(j) \propto 
    \begin{cases}
        \exp\left(\frac{j-J}{\Lambda}\right),\,&\text{if}\,J<J_{\max}
        \\
        \frac{1}{J_{\max}},\,&\text{if}\,J=J_{\max}
    \end{cases}
    \label{eq:level-weights}
\end{equation}
with 
\begin{equation}
    \sum_{j=1}^Jp_J(j) = 1, \, \forall J
\end{equation}
and the back-tracking parameter $\Lambda\in\mathbb{R}^+$, controlling the width of $p_J(j)$.\\

\noindent
We construct the likelihood level $L^{(j)}_{\{\mathcal{M}^k\}_{\mathcal{I}^k}}$ by alternately sampling the ladder-step $j$ according to the probability $p_J(j)$ using Metropolis-Hastings and sampling model-flux combinations $(i, v_{\mathcal{M}^k_i})$ according to $p(i,v_{\mathcal{M}_i^k} \mid j)$ using RJMCMC.
This alternate sampling produces samples (indicated by superscript $(n)$) 
\begin{equation}
    j^{(n)} \sim p_J(j) \quad \text{and} \quad (i^{(n)}, v^{(n)}_{\mathcal{M}_i^k}) \sim p(i,v_{\mathcal{M}_i^k}|j^{(n)})
\end{equation}
Specifically, the likelihood level $L^{(j)}_{\{\mathcal{M}^k\}_{\mathcal{I}^k}}$ is created after sampling a set of samples with likelihood higher than the previous level $L^{(j-1)}_{\{\mathcal{M}^k\}_{\mathcal{I}^k}}$ and setting  $L^{(j)}_{\{\mathcal{M}^k\}_{\mathcal{I}^k}}$ to the $(1-\alpha)$-quantile of the likelihoods of the samples, where $\alpha \in (0,1)$ is the so-called ``compression factor''.
After the likelihood level $L^{(j)}_{\{\mathcal{M}^k\}_{\mathcal{I}^k}}$ has been constructed, the prior mass for that level $X^{(j)}_{\{\mathcal{M}^k\}_{\mathcal{I}^k}}$ is estimated according to nested sampling theory~\cite{Ashton2022}:
\begin{equation}
X^{(j)}_{\{\mathcal{M}^k_i\}_{\mathcal{I}^k}} =
        \alpha\cdot X^{(j-1)}_{\{\mathcal{M}^k_i\}_{\mathcal{I}^k}} 
\end{equation}

\noindent
Finally, with $N_{\max}$ samples and $J_{\max}$ levels generated, the trapezoidal rule is used to approximate the model evidence
\begin{equation}
    p(\mathcal{D} \mid \{\mathcal{M}^k\}_{\mathcal{I}^k}) 
    \approx
    \underbrace{\sum\limits_{n=1}^{N_{\max}} 
    \mathcal{L}_{{\{\mathcal{M}^k\}_{\mathcal{I}^k}}}
    ( i^{(n)}, v_{\mathcal{M}_i^k}^{(n)} )
    \cdot
    \Delta{\mathcal{X}^{(n)}_{\{\mathcal{M}^k\}_{\mathcal{I}^k}}}}_{\hat p_{\mathcal{D} \mid \{\mathcal{M}^k\}_{\mathcal{I}^k}}}
    \label{eq:trapezoidalrjmcmc}
\end{equation}
with $\Delta{\mathcal{X}^{(n)}_{\{\mathcal{M}^k\}_{\mathcal{I}^k}}}$ the trapezoidal interval width for the $n$-th sample.
Specifically, each augmented flux sample $(i^{(n)}, v^{(n)}_{\mathcal{M}_i^k})$ is assigned a prior-mass coordinate, denoted by $\mathcal{X}^{(n)}_{\{\mathcal{M}^k\}_{\mathcal{I}^k}}$, such that those with a higher likelihood are assigned a lower prior mass. If multiple samples evaluate to the same likelihood (i.e., ``likelihood plateaus''), assigning prior masses to samples requires additional information~\cite{fowlie2021nested, schittenhelm2021nestedsamplinglikelihoodplateaus}. Here we rely on so-called ``likelihood tie-breakers''~\cite{brewer2016dnest4}, which resolve the issue of likelihood plateaus~\citep{fowlie2021nested}. \\
\noindent
The prior-mass coordinate $\mathcal{X}^{(n)}_{\{\mathcal{M}^k\}_{\mathcal{I}^k}}$ is estimated by sampling uniformly from the interval $I_{j^{(n)}}$ given by
\begin{equation}
    I_{j^{(n)}} = 
    \begin{cases}
        \left( X_{\{\mathcal{M}^k\}_{\mathcal{I}^k}}^{(j^{(n)}+1)}, \, 
        X_{\{\mathcal{M}^k\}_{\mathcal{I}^k}}^{(j^{(n)})} \right],&\text{if}\, j^{(n)}<J_{\max}
        \\
        \left[ 0, \, 
        X_{\{\mathcal{M}^k\}_{\mathcal{I}^k}}^{(j^{(n)})} \right]
        &\text{if}\, j^{(n)}=J_{\max}
        \end{cases}
\end{equation}

\noindent
With that, the trapezoidal interval width $\mathcal{X}^{(n)}_{\{\mathcal{M}^k\}_{\mathcal{I}^k}}$ for the $n$-th sample is given by
\begin{equation}
    \Delta{\mathcal{X}^{(n)}_{\{\mathcal{M}^k\}_{\mathcal{I}^k}}} =
    \frac{1}{2}
    \left(
        {\mathcal{X}^{(n_-)}_{\{\mathcal{M}^k\}_{\mathcal{I}^k}}}
        -
        {\mathcal{X}^{(n_+)}_{\{\mathcal{M}^k\}_{\mathcal{I}^k}}}
    \right)
\label{eq:sample-prior-weights}
\end{equation}
with
\begin{equation}
    {\mathcal{X}^{(n_-)}_{\{\mathcal{M}^k\}_{\mathcal{I}^k}}} = 
    \max\left(
    \{0\} \cup
    \left\{
        {\mathcal{X}^{(m)}_{\{\mathcal{M}^k\}_{\mathcal{I}^k}}} \mid
        m\in\{1,\cdots,N_{\max}\},\,
        {\mathcal{X}^{(m)}_{\{\mathcal{M}^k\}_{\mathcal{I}^k}}}  <
        {\mathcal{X}^{(n)}_{\{\mathcal{M}^k\}_{\mathcal{I}^k}}} 
    \right\}
    \right)
\end{equation}
and
\begin{equation}
    {\mathcal{X}^{(n_+)}_{\{\mathcal{M}^k\}_{\mathcal{I}^k}}} = 
    \min\left(
    \{1\} \cup
    \left\{
        {\mathcal{X}^{(m)}_{\{\mathcal{M}^k\}_{\mathcal{I}^k}}} \mid
        m\in\{1,\cdots,N_{\max}\},\,
        {\mathcal{X}^{(m)}_{\{\mathcal{M}^k\}_{\mathcal{I}^k}}}  >
        {\mathcal{X}^{(n)}_{\{\mathcal{M}^k\}_{\mathcal{I}^k}}} 
    \right\}
    \right)
\end{equation}
where the $0$ and $1$ are added to the sets to deal with the two samples that have only one neighboring sample in prior mass space.

\subsubsection{Posterior model set-averaged flux samples}

With the estimated evidence in Eq.~\eqref{eq:trapezoidalrjmcmc}, normalized posterior weights $\eta^{(n)}\in[0,1]$ are computed for each sample $(i^{(n)},v_{\mathcal{M}_i^k}^{(n)})$
\begin{equation}
  \eta^{(n)}  = 
  \frac{ \mathcal{L}_{{\{\mathcal{M}^k\}_{\mathcal{I}^k}}}
    (i^{(n)},v_{\mathcal{M}_i^k}^{(n)})
  \cdot
    \Delta{\mathcal{X}^{(n)}_{\{\mathcal{M}^k\}_{\mathcal{I}^k}}}
  }
  {\hat{p}_{\mathcal{D}\mid \{\mathcal{M}^k\}_{\mathcal{I}^k}}}
  \label{eq:posterior-weights}
\end{equation}

\noindent
To obtain the final $T$ posterior augmented flux samples, the samples $(i^{(n)},{v_{\mathcal{M}_i^k}}^{(n)})$ are resampled according to their posterior weights $\eta^{(n)}$, such that the relative frequency of each point in the resampled set approximately represents its posterior weight. 
The model set-averaged fluxes $v_{\{\mathcal{M}^k\}_{\mathcal{I}^k}}$ are obtained by marginalizing over the $i$-component of the posterior augmented flux samples, where the $i$-component represents the model index of the posterior augmented flux samples. 
Furthermore, by counting the samples that point to a specific model $\mathcal{M}_i^k$, the approximate posterior probability of this model is obtained given the considered model set.
In the main text, for this we use the notation 
\begin{equation}
    p(\mathcal{M}_{i}^{k} | \{\mathcal{M}^k\}_{\mathcal{I}^k}, \mathcal{D}) \approx \frac{\# \mbox{samples} (\mathcal{M}_{i}^{k})}{\# \mbox{samples} (\{ \mathcal{M}^{k}\}_{\mathcal{I}^{k}})}, \ \forall i, k
\end{equation}
where $\# \mbox{samples} (\mathcal{M}_{i}^{k})$ denotes how often a model $\mathcal{M}_i^k$ was sampled and
$\# \mbox{samples} (\{ \mathcal{M}^{k}\}_{\mathcal{I}^{k}})$ is the total number of posterior samples drawn for the model set $k$.

\filbreak
\noindent\begin{minipage}{\textwidth}

\subsubsection{The TDNS for \AC{} algorithm}

\begin{algorithm}[H]
\caption{TDNS for \AC}
\label{alg:TDNS}
\begin{algorithmic}
\State \textbf{Input:} 
\State Model set $\{\mathcal{M}^k\}_{\mathcal{I}^k}$
\State Priors for fluxes $p(v_{\mathcal{M}_i^k} \mid \mathcal{M}_i^k)$ and all models constituting the model set $p(\mathcal{M}_i^k \mid \{\mathcal{M}^k\}_{\mathcal{I}^k})$
\State Maximum number of samples to draw $N_{\max}$
\State Maximum number of likelihood levels $J_{\max}$
\State Level interval size $l_{new}$
\State Back-tracking parameter $\Lambda$
\State Compression factor $\alpha\in(0,1)$
\State \textbf{Output:} 
\State Posterior model set-averaged flux samples $\{v_{\{\mathcal{M}^k\}_{\mathcal{I}^k}}^{(t)}\}_{t=1}^{T} \, \text{ with } \, v_{\{\mathcal{M}^k\}_{\mathcal{I}^k}}
^{(t)} \sim \, p(v_{\{\mathcal{M}^k\}_{\mathcal{I}^k}} \mid \mathcal{D})$ 
\State Evidence estimate $\hat{p}_{\mathcal{D} \mid \{\mathcal{M}^k\}_{\mathcal{I}^k}}$
\State \textbf{Procedure:}
\State $n\gets0$
\State $J\gets 1$ \Comment{Set likelihood level counter}
\State $L^{(J)}_{\{\mathcal{M}^k\}_{\mathcal{I}^k}} \gets 0$ \Comment{Create initial likelihood level}
\State $
X^{(J)}_{\{\mathcal{M}^k\}_{\mathcal{I}^k}} \gets 1$ \Comment{Set initial prior mass}
    \While{$J \leq J_{\max}$ }
        \State $\mathbf{L}_{\text{seen}}\gets \emptyset$ \Comment{Initialize store for seen likelihoods}
        \State $l\gets 0$
        \While{$l < l_{new}$}
            \State $n \gets n+1$ \Comment{Increment number of samples}
            \State 
            Metropolis-Hastings sample $j^{(n)}\sim p_J(j)$
            \Comment{Level move; cf.~Eq.~\eqref{eq:level-weights}}
            \State
            RJMCMC sample $(i^{(n)}, v^{(n)}_{\mathcal{M}^k_{i}}) \sim p(i, v_{\mathcal{M}^k_{i}}\mid j^{(n)})$
            \Comment{Parameter move; cf.~Eq.~\eqref{eq:likelihood-constrained-priors}}
            \If{$n=N_{\max}$}
                \State Go to \textbf{Evaluation} 
            \EndIf
            \If{$\mathcal{L}_{\{\mathcal{M}^k\}_{\mathcal{I}^k}}(i^{(n)}, v^{(n)}_{\mathcal{M}^k_i})
            > L^{(J)}_{\{\mathcal{M}^k\}_{\mathcal{I}^k}}$}
               \State $l \gets l+1$ \Comment{Increment number of likelihoods in next level interval}
               \State $\mathbf{L}_{\text{seen}} \gets \mathbf{L}_{\text{seen}} \cup 
            \{
            \mathcal{L}_{\{\mathcal{M}^k\}_{\mathcal{I}^k}}(i^{(n)}, v^{(n)}_{\mathcal{M}^k_i})
            \}$ \Comment{Store likelihood of $n$-th sample}
            \EndIf
        \EndWhile
        \State $\triangleright$ Create next likelihood level based on seen likelihoods
        \State $J \gets J+1$ \Comment{Increment likelihood level counter}
        \State $L^{(J)}_{\{\mathcal{M}^k_i\}_{\mathcal{I}^k}} \gets
        \operatorname{Quantile}(\mathbf{L}_{\text{seen}},\,1-\alpha) $
        \State  
        \State $
        X^{(J)}_{\{\mathcal{M}^k\}_{\mathcal{I}^k}} \gets
        \alpha\cdot X^{(J-1)}_{\{\mathcal{M}^k\}_{\mathcal{I}^k}} 
        $ \Comment{Set prior mass by compression of forerunner level}
    \EndWhile
    \While{$n < N_{max}$} 
            \State $n \gets n+1$ \Comment{Increment number of samples}
            \State 
            Metropolis-Hastings sample $j^{(n)}\sim p_J(j)$
            \Comment{Level move; see Eq.~\eqref{eq:level-weights}}
            \State
            RJMCMC sample $(i^{(n)}, v^{(n)}_{\mathcal{M}^k_{i}}) \sim p(i, v_{\mathcal{M}^k_{i}}\mid j^{(n)})$
            \Comment{Parameter move; cf.~Eq.~\eqref{eq:likelihood-constrained-priors}}
    \EndWhile
\State \textbf{Evaluation:} 
\State 1) Estimate the prior-mass interval widths $\{\Delta\mathcal{X}_{\{\mathcal{M}^k\}_{\mathcal{I}^k}}^{(n)}\}_{n=1}^{N_{\max}}$  
according to Eq.~\eqref{eq:sample-prior-weights}
\State 2) Use
$\{\Delta\mathcal{X}_{\{\mathcal{M}^k\}_{\mathcal{I}^k}}^{(n)}\}_{n=1}^{N_{\max}}$  
to obtain evidence estimate 
$\hat{p}_{\mathcal{D} \mid \{\mathcal{M}^k\}_{\mathcal{I}^k}}$
according to Eq.~\eqref{eq:trapezoidalrjmcmc}
\State 3) 
Reweigh augmented flux space samples $\{(i^{(n)}, v^{(n)}_{\mathcal{M}_i^k})\}_{n=1}^{N_{\max}}$ according to Eq.~\eqref{eq:posterior-weights}
and marginalize out $i$-component so that $T$  model averaged flux posterior samples $\{v_{\{\mathcal{M}^k\}_{\mathcal{I}^k}}^{(t)}\}_{t=1}^{T}$ are obtained 
\end{algorithmic}
\end{algorithm}
\end{minipage}

\FloatBarrier
\clearpage

\subsubsection{Parameterization, diagnostics, and reproducibility}

TDNS for \AC{} intakes a set of parameters that control the sampling processes, allowing the algorithm to be adapted to the difficulty of the given \cmfa{} problem. \\
\noindent
The problem-specific parameters are $J_{\max}$, $l_{new}$, and $\Lambda$. Suitable values for these parameters are determined using tuning runs. Generally, higher values of these three parameters provide more robust results, but also increase computational costs.
We found that more levels are required when the posterior is concentrated in a small region compared to the prior. This finding has also been reported previously~\citep{Hu2024}.
Higher values of $\Lambda$ and $l_{new}$ are beneficial, when the likelihood has pronounced multi-modality and the MCMC sampling is slow to explore the augmented flux space, see for example~\cite{brewer2016dnest4}. \\
Two further parameters are less closely tied to the problem characteristic in the sense, that they do not require tuning runs: the compression factor $\alpha$ and the maximum number of samples to be drawn, $N_{\max}$. The default compression factor is set to $\alpha=e^{-1}$~\cite{brewer2016dnest4}. The $N_{\max}$ parameter is flexible and adaptable because sampling runs can always be continued if more samples are required. For example, it is necessary to continue sampling if the desired $J_{\max}$ levels have not yet been reached or if the information contained by the $T$ posterior samples should be further increased. The ability to extend runs is a property that sets DNS and TDNS apart from other nested sampling algorithms. 
Notably, problems with similar concentration of posterior mass relative to the prior exhibit comparable computational cost~\cite{Hu2024}, enabling transfer of TDNS parameterizations across related inference tasks and thereby reducing the need for extensive tuning. We have previously tested TDNS configurations for different data sets, providing reference points for tuning \AC{}~\cite{jadebeck2025}.\\
\noindent 
All parameter values used for all case studies are found in the respective computational scripts.
The specific parameter values, which we used in our \textit{E.~coli} case studies are also reported in \Cref{sec:TDNS-diagnostics}. 

For diagnostics, we followed the recommendation in \cite{brewer2016dnest4} and checked that the increase in likelihood values for decreasing $X_{\{\mathcal{M}^k\}_{\mathcal{I}^k}}$ slowed down, resulting in a well-visible peak in the associated posterior weights of the samples (Eq.~\eqref{eq:posterior-weights}) as a function of $X_{\{\mathcal{M}^k\}_{\mathcal{I}^k}}$; see~\cite{brewer2016dnest4} for more details and ~\Cref{sec:TDNS-diagnostics}.

We recommend to rerun TDNS for \AC{} several times with different random seeds for new problems.
Since there are no explicit starting points in TDNS (uniform sampling, i.e., prior sampling mixes extremely well on convex polytopes~\cite{Jadebeck2023}), changing the random seed ensures that the algorithm uses different samples to explore flux and model spaces. 
In addition, multiple runs improve evidence estimation.

\FloatBarrier
\clearpage
\section{\, \itshape Triangulus\normalfont\textbf{: Additional information}}
Sources and scripts to replicate results are available at \href{https://github.com/JuBiotech/Supplement-to-Jadebeck-et-al.-2026}{https://github.com/JuBiotech/Supplement-to-Jadebeck-et-al.-2026}.

\subsection{Prior specification}

\begin{table}[h!]
    \centering
    \caption{\textbf{Model and flux priors for the \tri{} study.}
    As the DOFs increase, the flux priors are effectively diffused across the larger flux spaces.}
    \begin{tabular}{ccc|cccc}
        \toprule
        $k$ & $i$ & DOF & $p(\{\mathcal{M}^k\}_{\mathcal{I}^k})$ & $p(\mathcal{M}_i^k\mid \{\mathcal{M}^k\}_{\mathcal{I}^k})$ & $p(\mathcal{M}^k_i)$ & $p(v\mid \mathcal{M}^k_i)$ \\ 
        \midrule
$1$ & $1$ & 1 & $1/3$ & $1/2$ & $1/6$ & $\begin{cases}1/1000, & \text{if } v\in\mathcal{P}_{\mathcal{M}^1_1} \\ 0, & \text{otherwise} \end{cases}$ \\
\hline
$1$ & $2$ & 2 & & $1/2$ & $1/6$ &  $\begin{cases}1/1000^2, & \text{if } v\in\mathcal{P}_{\mathcal{M}^1_2} \\ 0, & \text{otherwise} \end{cases}$ \\
\hline
$2$ & $1$ & 2 & $1/3$ & $1/2$ & $1/6$ &  $\begin{cases}2/1000^2, & \text{if } v\in\mathcal{P}_{\mathcal{M}^2_1} \\ 0, & \text{otherwise} \end{cases}$ \\
\hline
$2$ & $2$ & 3 & & $1/2$ & $1/6$ &  $\begin{cases}2/1000^3, & \text{if } v\in\mathcal{P}_{\mathcal{M}^2_2} \\ 0, & \text{otherwise} \end{cases}$\\ 
\hline
$3$ & $1$ & 3 & $1/3$ & $1/2$ & $1/6$ &  $\begin{cases}2/(5\cdot1000^2), & \text{if } v\in\mathcal{P}_{\mathcal{M}^3_1} \\ 0, & \text{otherwise} \end{cases}$\\ 
\hline
$3$ & $2$ & 4 & & $1/2$ & $1/6$ &  $\begin{cases}2/(5\cdot1000^3), & \text{if } v\in\mathcal{P}_{\mathcal{M}^3_2} \\ 0, & \text{otherwise} \end{cases}$\\ 
        \bottomrule
    \end{tabular}
    \label{tab:priors_triangulus}
\end{table}

\FloatBarrier

 \begin{figure}[ht!]
    \centering
    \includegraphics[width=1\linewidth]{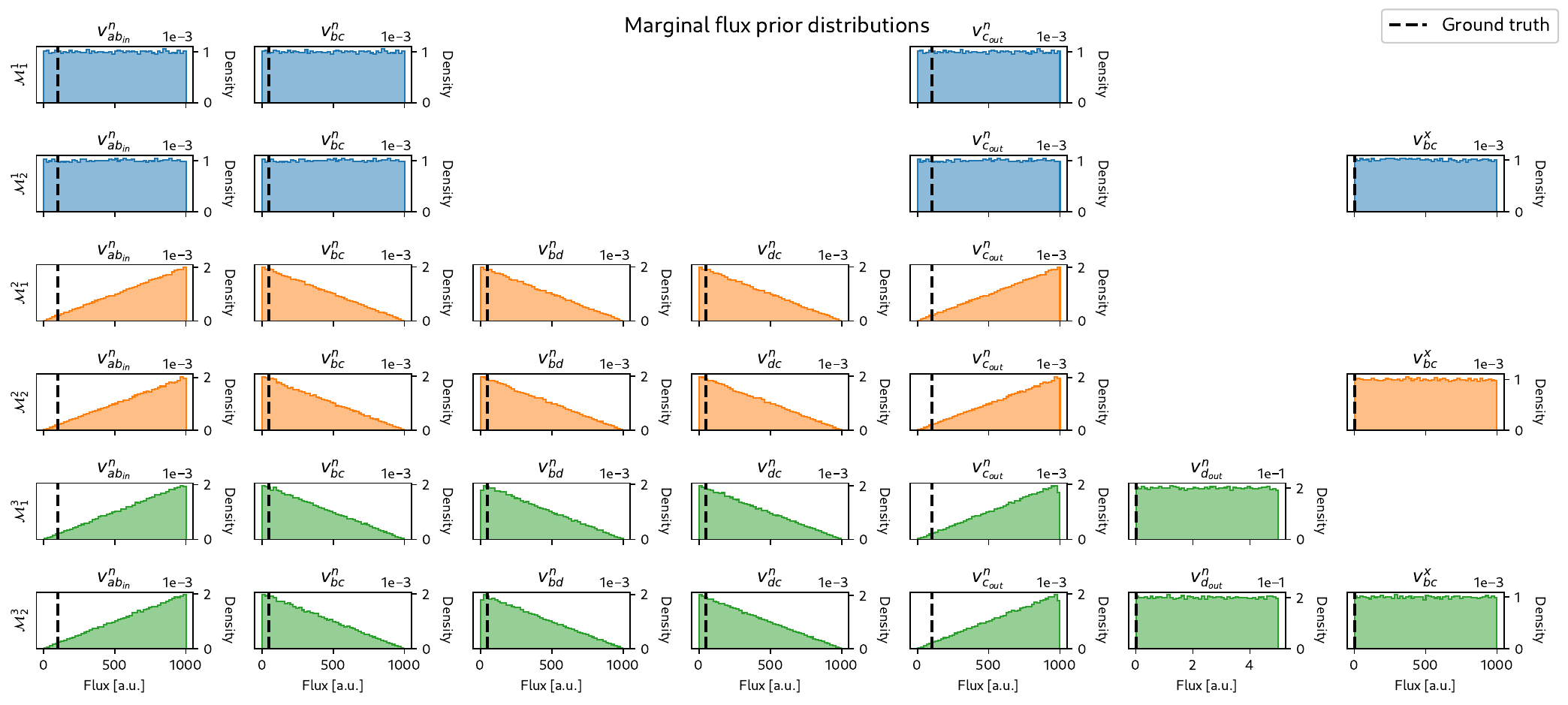}
    \caption{\textbf{Marginal flux priors for the \tri{} case study.} 
    Each row corresponds to a different model and shows all its fluxes.
    $\mathcal{M}_2^1$ is the data generating \tri{} model with 
    ground truth fluxes indicated by dashed lines.
    For the models of the green model set, zero values of \netflux{dout} and \xchflux{bc} collapse the models to the (nested) \tri{} model.}
    \label{fig:triangulus_priors} 
\end{figure}

\FloatBarrier
\newpage

\subsection{Marginal posteriors}

\begin{figure}[ht!]
    \centering
    \includegraphics[width=1\linewidth]{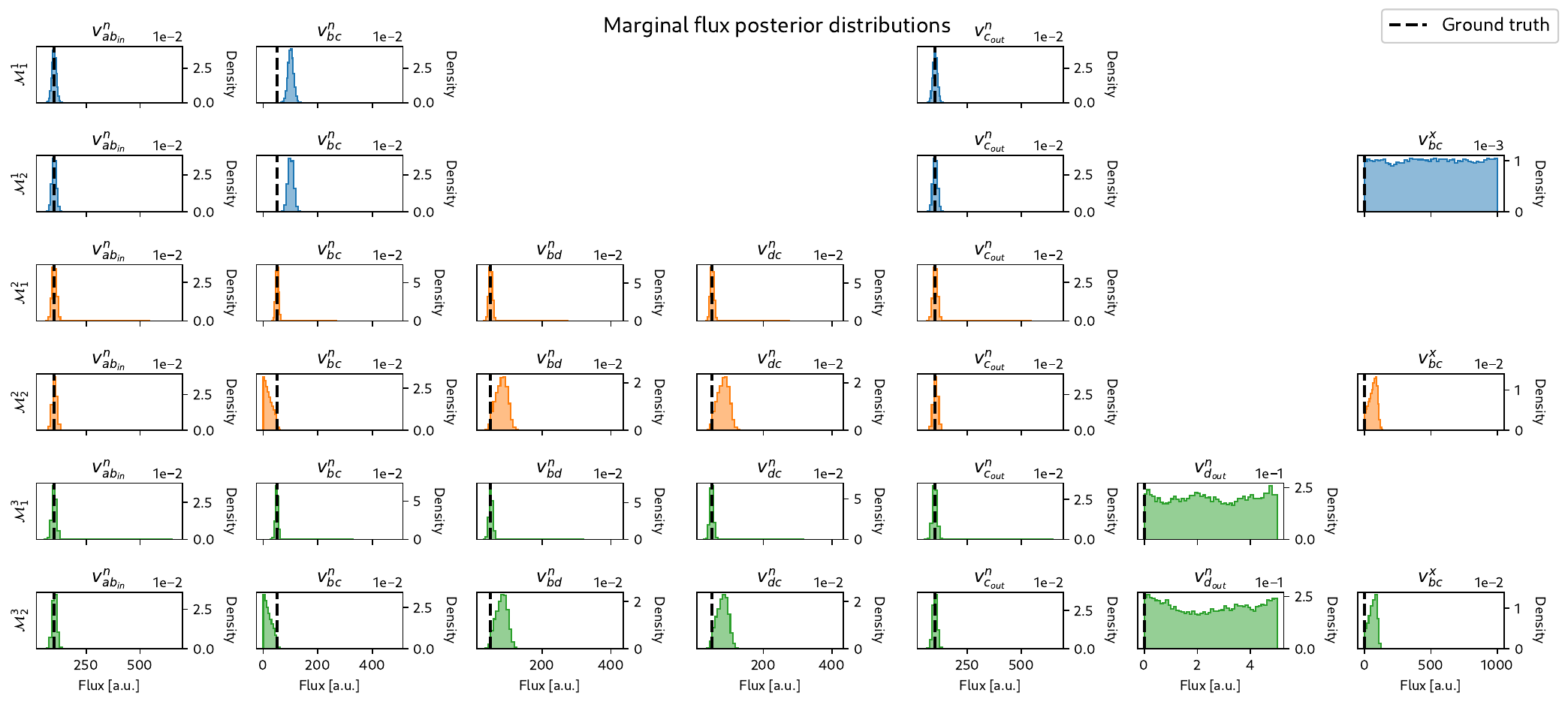}
    \caption{\textbf{Single-model flux posteriors for the \tri{} case study.}
    Each row corresponds to a different model and shows all of its fluxes. The ground truth fluxes are indicated by dashed lines. Only the ground truth model ($\mathcal{M}^2_1$, third row) accurately recovers all the fluxes. The more complex models exhibit biases for the internal fluxes. The models in the blue model set are unable to explain the data due to the way they are constructed.}
    \label{fig:triangulus_posteriors} 
\end{figure}

\subsection{TDNS diagnostics}

\begin{figure}[ht!]
    \centering
    \includegraphics[width=0.85\linewidth]{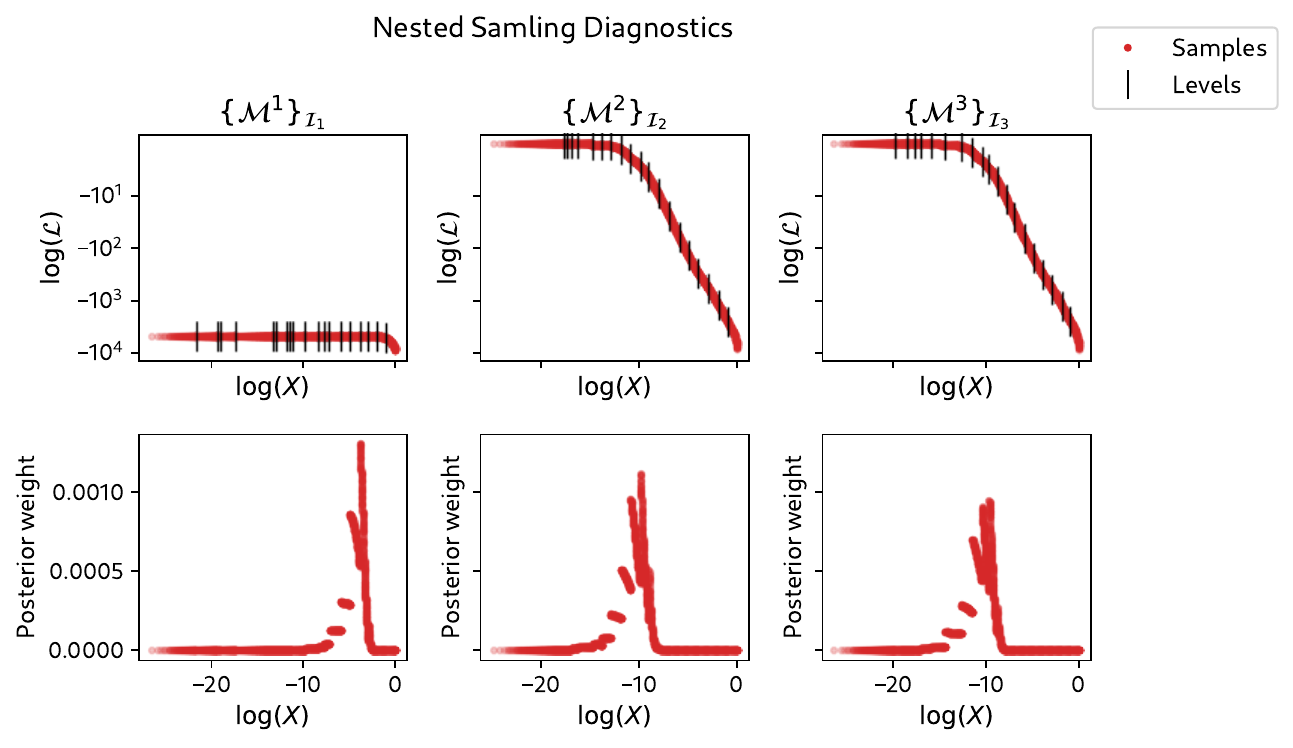}
    \caption{\textbf{Diagnostics of the TDNS runs for the \tri{} case study. Each column corresponds to one model subset.}
    The first row shows the log-likelihood over the prior mass $X_{\{ \mathcal{M}^k \}_{\mathcal{I}^k} }$. The increase in likelihood levels off once sufficient prior mass has been accumulated. 
    In the bottom row, sharp peaks in the posterior weight indicate that the typical set has been identified. There is no posterior weight for very small prior masses $\log(X_{\{ \mathcal{M}^k \}_{\mathcal{I}^k} })$.
    \AC{} took 30.7~seconds on a laptop (Intel Core i7-13700H) with each model set  $\{ \mathcal{M}^k \}_{\mathcal{I}^k}$ run sequentially.} 
    \label{fig:triangulus_convergence} 
\end{figure}

\FloatBarrier
\clearpage

\section{\, \itshape \ECfull \normalfont\textbf{use case: Additional information}}
Sources and scripts to replicate results are available at \href{https://github.com/JuBiotech/Supplement-to-Jadebeck-et-al.-2026}{https://github.com/JuBiotech/Supplement-to-Jadebeck-et-al.-2026}.
\subsection{Reference model and flux map}

\begin{figure}[ht!]
    \centering
    \includegraphics[width=0.6\linewidth]{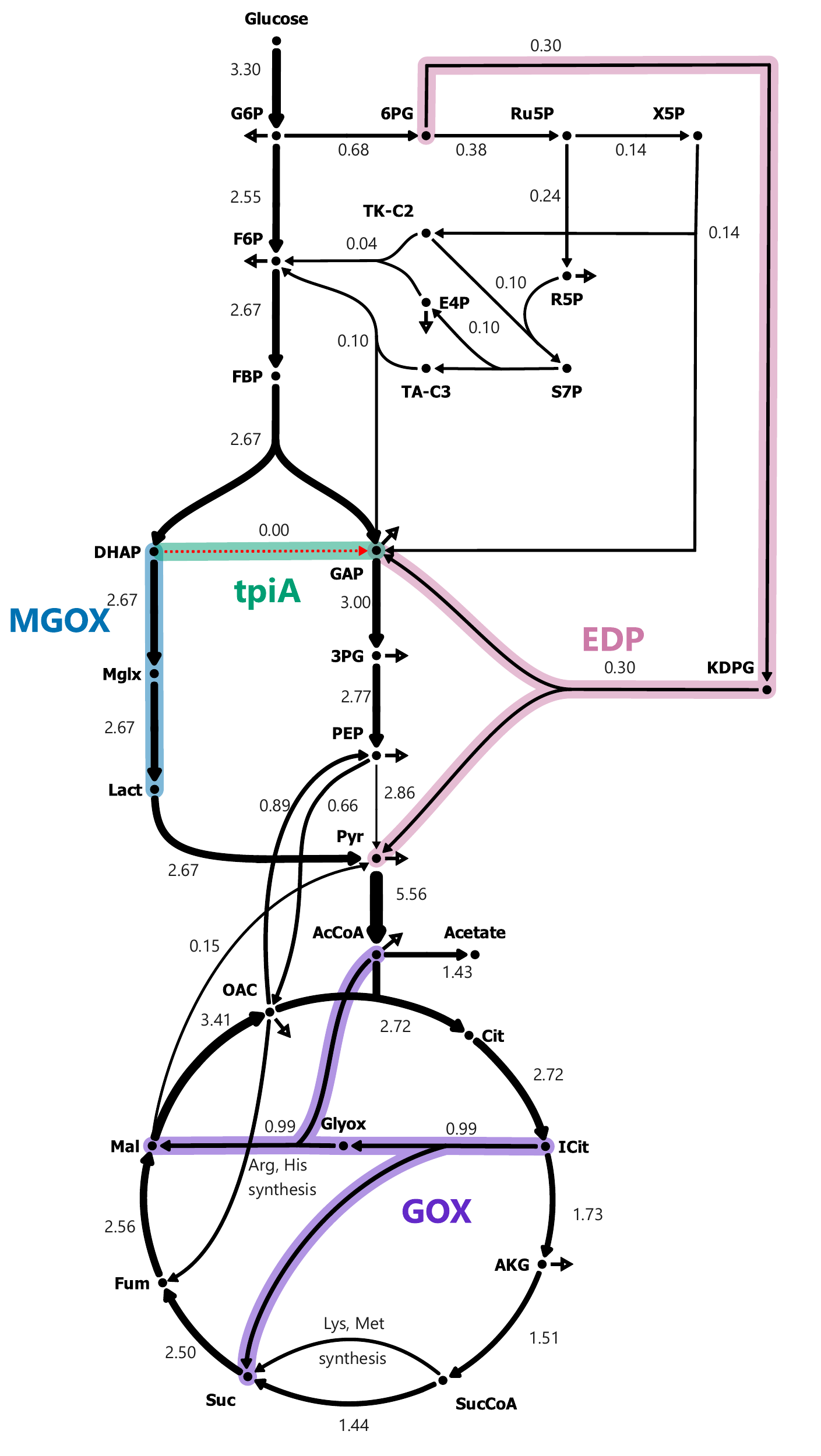}
    \caption{\textbf{Reference metabolic map for the \ECfull{} case study.}
    The latent pathways with uncertain metabolic activity (GOX - glyoxylate shunt, MGOX - methylglyoxal pathway, EDP - Entner-Doudoroff pathway), as well as the TPI pathway with the triosephosphate isomerase gene knocked out, are highlighted. Values give net fluxes in \SI{}{mmol.g_{CDW}^{-1}.h^{-1}}. The reference model has 11 (independent) net fluxes and 23 exchange fluxes, resulting in 34 DOF.}
    \label{fig:model}
\end{figure}

\FloatBarrier
\clearpage

\subsection{Model subsets}

\begin{table}[h!]
    \centering
    \caption{\textbf{Model subsets for the \ECfull{} case study.}
    Out of the 2\textsuperscript{4}=16 combinations of the four latent pathways with uncertain metabolic activity, four  combinations are stoichiometrically impossible because they contain a dead-end metabolite (i.e. DHAP), which renders either TPI, MGOX or both active. 
    Therefore, twelve stoichiometrically model sets are valid.
    The smallest model set $\{\mathcal{M}^9\}_{\mathcal{I}^9}$ (TPI) contains \num{268435456} unique model structures. The most comprehensive set $\{\mathcal{M}^6\}_{\mathcal{I}^6}$ contains \num{17179869184} different model variants and is 64 times larger. 
    Together, all model sets contain \num{46976204800} models
    The DOF range over all models is \num{9}-\num{46}.
    Each model subset is based on a distinct reaction set, specified in a single FluxML file.}
    \begin{tabular}{cc|ccc}
        \toprule
        $k$ & reactions sets & DOF range & cardinality & FluxML filename\\
        \midrule
1 & EDP-GOX-MGOX & 11-44 & 2\textsuperscript{33} &
\texttt{e\_coli\_+EDP+GOX+MGOX-TPI.fml} 
\\
\hline
2 & GOX-MGOX & 10-41 & 2\textsuperscript{31} & 
\texttt{e\_coli\_-EDP+GOX+MGOX-TPI.fml} 
\\
\hline
3 & GOX-TPI & 10-40 & 2\textsuperscript{30} & 
\texttt{e\_coli\_-EDP+GOX-MGOX+TPI.fml} 
\\
\hline
4 & EDP-GOX-TPI & 11-43 & 2\textsuperscript{32} &
\texttt{e\_coli\_+EDP+GOX-MGOX+TPI.fml} 
\\
\hline
5 & GOX-MGOX-TPI & 11-43 & 2\textsuperscript{32}  &
\texttt{e\_coli\_-EDP+GOX+MGOX+TPI.fml} 
\\
\hline
6 & EDP-GOX-MGOX-TPI & 12-46 & 2\textsuperscript{34}  &
\texttt{e\_coli\_+EDP+GOX+MGOX+TPI.fml} 
\\
\hline
7 & EDP-MGOX & 10-41 & 2\textsuperscript{31}  &
\texttt{e\_coli\_+EDP-GOX+MGOX-TPI.fml} 
\\
\hline
8 & MGOX & 9-38 & 2\textsuperscript{29}  &
\texttt{e\_coli\_-EDP-GOX+MGOX-TPI.fml} 
\\
\hline
9 & TPI & 9-37 & 2\textsuperscript{28}  &
\texttt{e\_coli\_-EDP-GOX-MGOX+TPI.fml} 
\\
\hline
10 & EDP-TPI & 10-40 & 2\textsuperscript{30}  &
\texttt{e\_coli\_+EDP-GOX-MGOX+TPI.fml} 
\\
\hline
11 & MGOX-TPI & 10-40 & 2\textsuperscript{30}  &
\texttt{e\_coli\_-EDP-GOX+MGOX+TPI.fml} 
\\
\hline
12 & EDP-MGOX-TPI & 11-43 & 2\textsuperscript{32}  &
\texttt{e\_coli\_+EDP-GOX+MGOX+TPI.fml} 
\\
        \bottomrule
    \end{tabular}
    \label{tab:modelsubsets_ecoli}
\end{table}

\subsection{Prior specification}
\label{ssec:priorspec}

For our studies, we use the following setting for the models
\begin{equation}
    p(\{\mathcal{M}^k\}_{\mathcal{I}^k})=\tfrac{1}{12} \, \text{ and } \, p(\mathcal{M}_i^k\mid \{\mathcal{M}^k\}_{\mathcal{I}^k})=\tfrac{1}{|\{\mathcal{M}^k\}_{\mathcal{I}^k}\}|}.
\end{equation}
With that, the model priors are 
\begin{equation}
    p(\mathcal{M}^k_i)=\tfrac{1}{12\cdot |\{\mathcal{M}^k\}_{\mathcal{I}^k}|}
\end{equation}
For the fluxes, we use a polytope-constrained uniform prior with additional boxed constraints, i.e. \num{100} times the uptake rate in both directions
\begin{equation}
    \text{\SI{-330}{\milli \mole \per \gCDW \per hour}} 
    \leq v_{\mathcal{M}^k_{i}} \leq  
    \text{\SI{330}{\milli \mole \per \gCDW \per hour}},
    \,\forall i\in\mathcal{I}^k
\end{equation}
where the vector inequality is to be understood component-wise.

\FloatBarrier
\clearpage

\subsection{Marginal flux posteriors}

\begin{figure}[ht!]
    \centering
    \includegraphics[width=0.94\linewidth]{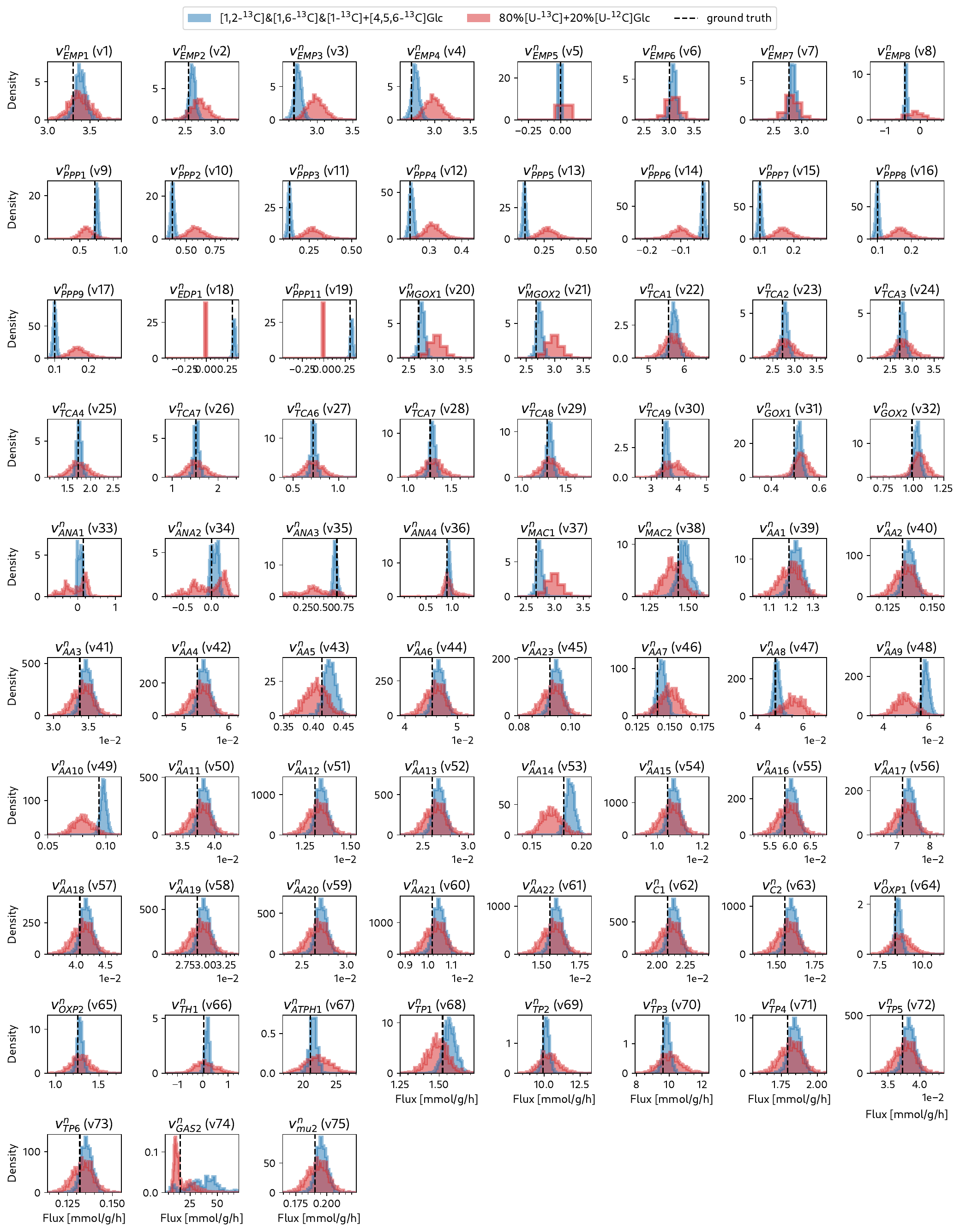}
    \caption{\textbf{Marginal full posteriors for all net fluxes contained in at least one of the models in the full model set $\{\mathcal{M}^k\}_\mathcal{K}$.} Posteriors are shown for both the single-ILE evaluation and the multi-ILE evaluation. As expected, the marginal distributions are wider for the single dataset evaluation.
    The names of the fluxes refer to the naming convention used in the FluxML files, while the names in brackets refer to the original study in \cite{Long2019}.}
    \label{fig:overview}
\end{figure}

\clearpage
\FloatBarrier

\begin{figure}[ht!]
    \centering
    \includegraphics[width=0.92\linewidth]{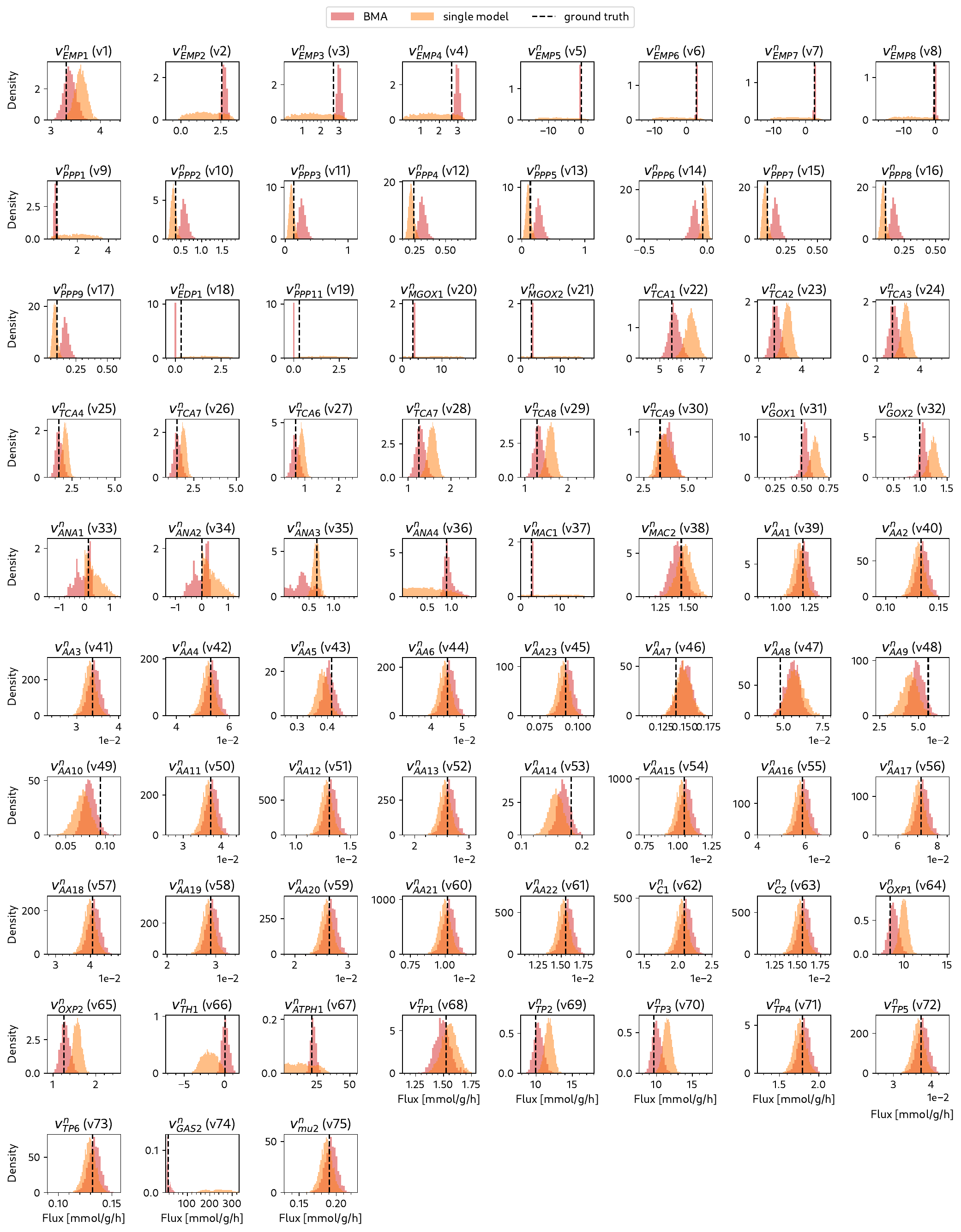}
    \caption{\textbf{Comparison of BMA-based inference using \AC{} and the traditional single-model inference approach for the single-ILE evaluation.}
    The marginal posteriors for all net fluxes are shown. For the traditional analysis, a super-model based on the original study by \cite{Long2019} was used. This model contains all latent pathways (EDP-GOX-MGOX-TPI) with 38 DOF (12 independent net fluxes and 26 exchange fluxes). 
    Flux names refer to the naming convention used in the FluxML files, with names in brackets referring to those in the original study.
    Interestingly, the flux marginals for core fluxes in EMP and PPP are wider than for the single-model analysis compared to the BMA approach based on an extensive model set, indicating more uncertainty. This discrepancy arises because the data support a set of simpler models compared to the original model.
    }
    \label{fig:singleA}
\end{figure}

\clearpage
\FloatBarrier

\begin{figure}[ht!]
    \centering
    \includegraphics[width=0.9\linewidth]{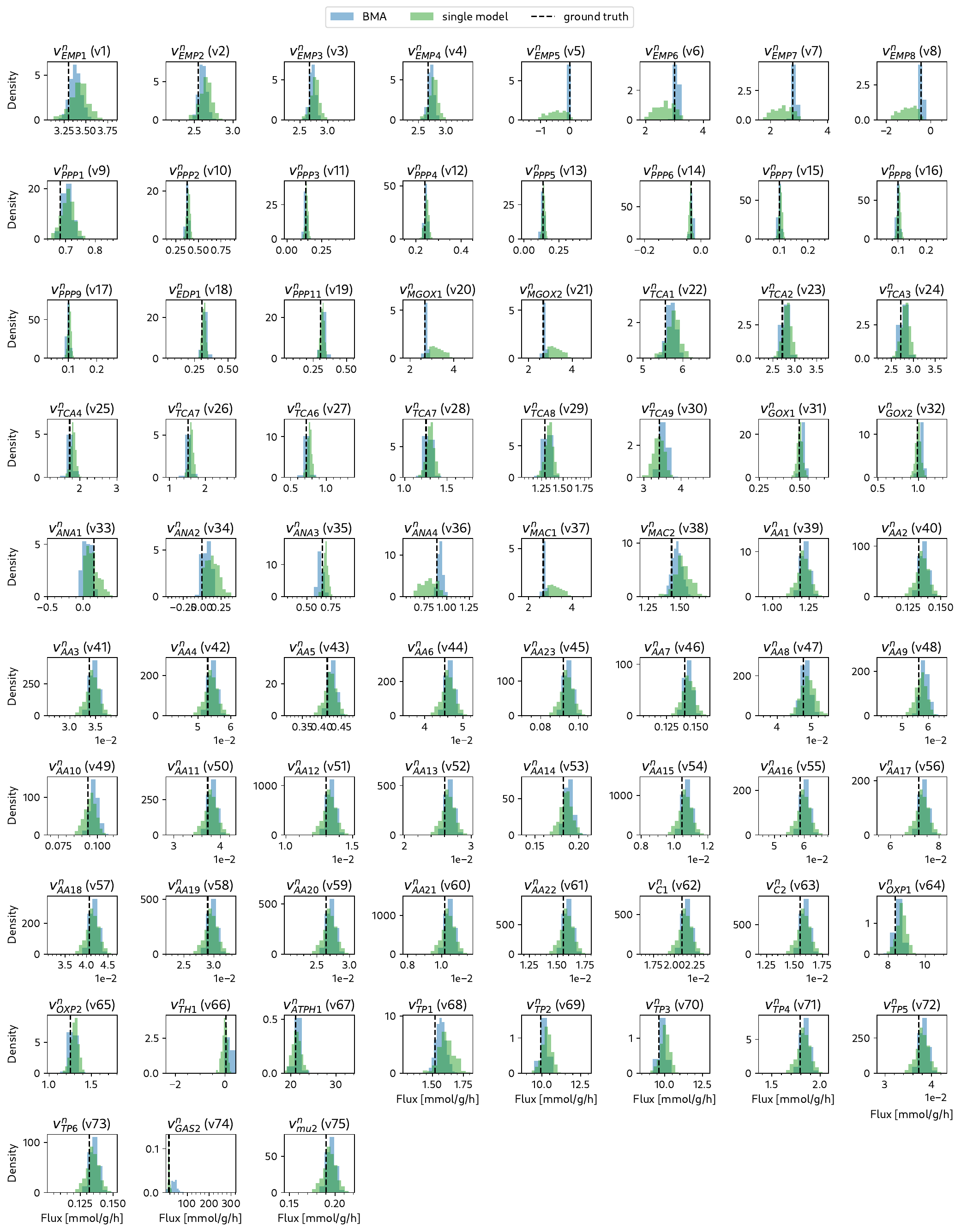}
    \caption{\textbf{Comparison of BMA-based inference generated using \AC{} and the single-model inference approach for the multi-ILE evaluation.}
    The marginal posteriors for all net fluxes are shown. For the single-model analysis, a super-model from the original study by \cite{Long2019} was used. This model contains all latent pathways (EDP-GOX-MGOX-TPI) and has 38 DOF (12 independent net fluxes and 26 exchange fluxes). Flux names refer to the naming convention used in the FluxML files, with names in brackets referring to those in the original study. Interestingly, the flux marginals for core fluxes in EMP and PPP are wider for the single-model analysis than for the BMA approach based on an extensive model set, indicating more uncertainty. This discrepancy is explained by the fact that BMA inference supports simpler models (effective model set), given the data, as compared to the original model. 
    Notably, the differences between single-model and multi-model approaches are starkly reduced compared to the results of the single-ILE dataset evaluation shown in \Cref{fig:singleA}.}
    \label{fig:BMA-BMSA}
\end{figure}

\clearpage
\FloatBarrier

\subsection{TDNS for \AC{} parameters \& diagnostics}
\label{sec:TDNS-diagnostics}

In our experiments, we used the following TDNS for \AC{} parameters for the single-ILE dataset evaluation
\begin{itemize}
    \item[-] $J_{\max} =$ \num{225}
    \item[-] $\Lambda =$ \num{50}
\end{itemize}
and for the multi ILE dataset evaluation
\begin{itemize}
    \item[-] $J_{\max} =$ \num{375}
    \item[-] $\Lambda =$ \num{80}
\end{itemize}
In both cases, $l_{new}$ was set to \num{360000} and \num{128} parallel RJMCMC chains were used.

\begin{figure}[ht!]
    \centering
    \rotatebox{90}{
        \begin{minipage}{0.9\textheight}
            \centering
            \includegraphics[width=1.0\linewidth]{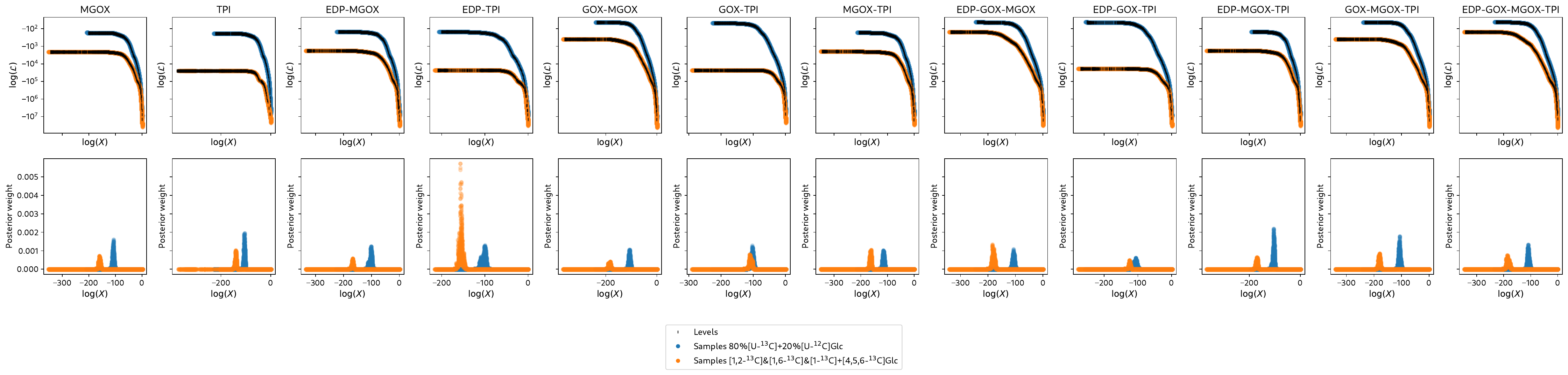}
            \caption{\textbf{Diagnostics of the TDNS for \AC{} runs for \ECfull{}.} 
            Each column corresponds to one of the model subsets in \Cref{tab:modelsubsets_ecoli}. In the upper row, the log-likelihood over the prior mass $X_{\{ \mathcal{M}^k \}_{\mathcal{I}^k} }$ is shown. The increase in likelihood levels off as soon as sufficient prior mass has been accumulated. In the lower row, sharp peaks in the posterior weight indicate that the typical set has been identified. Notably, there is no posterior weight for small prior masses. 
            The results of the single-ILE evaluation show a larger negative log-likelihood than the multi-ILE analysis, as the former contains only one-third of the measurements of the latter.}
        \end{minipage} 
    } \label{sifig:ecoli_convergence} 
\end{figure}

\clearpage
\FloatBarrier

\subsection{Reproducibility of results}
\label{ssec:reproducibility}

\begin{figure}[ht!]
    \centering
    \includegraphics[width=1.0\linewidth]{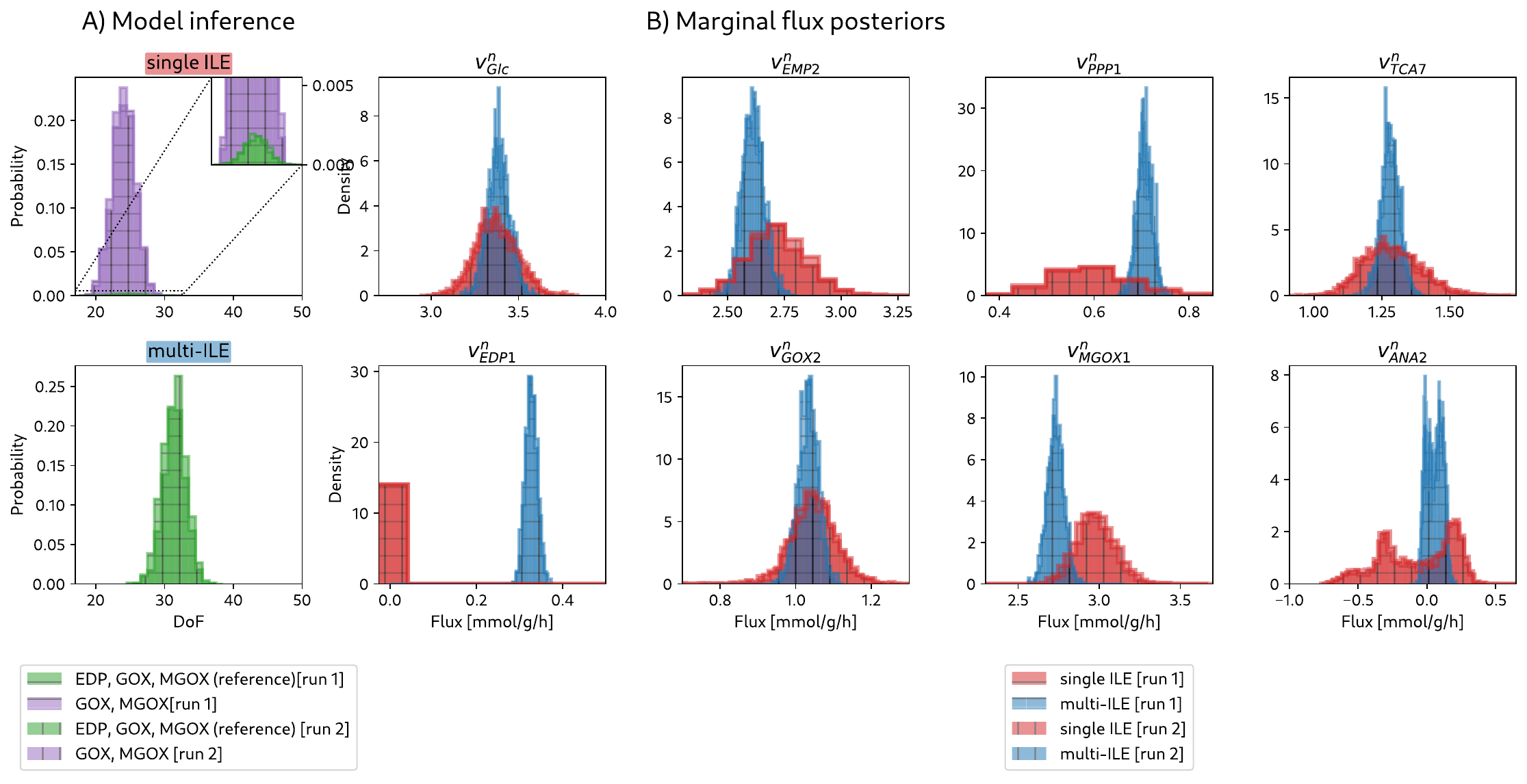}
    \caption{\textbf{Result summary for \AC{} run 1 and 2 of the single and multi-ILE evaluation.} 
    Hatches were added to distinguish the runs while keeping the colors. The results of both runs are the same.}
    \label{fig:reproducibility}
\end{figure}

To study the reproducibility of \AC{}, the inferences were re-run with a different random seed for both, the single-ILE and multi-ILE cases. 
Figure~\ref{fig:reproducibility} shows the results for the two runs with the same reference fluxes shown as in main Figure 2. The activation probabilities for the bidirectional reactions in both runs are given in ~\Cref{tab:reproducibility}. The two runs produced the same results, so we conclude that inferences computed by \AC{}  are reproducible.

\clearpage
\FloatBarrier

\begin{table}[htbp]
    \centering
    \caption{\textbf{Activation probability for the bidirectionality of all reactions for two single-ILE and two multi-ILE \AC{} runs.} 
    The maximum discrepancy in probability for the reaction bidirectionalities between runs for the single-ILE evaluation is around \SI{0.089} (TCA4\_v25) and around \SI{0.166} (TCA4\_v26) for the multi-ILE case. Conversely, many probabilities are identical between runs. Overall, this result is within the expected variance of TDNS (particularly the RJMCMC part), and we can therefore conclude that our runs are reproducible with respect to the fluxes.
    The model set inferred for the single-ILE evaluation does not contain 
    the EDP pathway, meaning we assign the value \num{0} to the activation probabilities to the bidirectionality of the EDP reactions PPP10\_v18 and PPP11\_v19.}
    \begin{tabular}{lrrrr}
    \toprule
     & Single ILE run1 & Single ILE run2 & Multi-ILE run1 & Multi-ILE run2 \\
     & Act. prob. & Act. prob. & Act. prob. & Act. prob. \\
    Reaction &  &  &  &  \\
    \midrule
    AA8\_v47 & 0.0000 & 0.0000 & 1.0000 & 1.0000 \\
    AA9\_v48 & 0.0000 & 0.0000 & 0.0000 & 0.0000 \\
    ANA1\_v33 & 0.0023 & 0.0000 & 0.0000 & 0.0000 \\
    ANA2\_v34 & 0.0000 & 0.0004 & 0.0000 & 0.0000 \\
    EMP2\_v2 & 1.0000 & 1.0000 & 1.0000 & 1.0000 \\
    EMP4\_v4 & 0.5170 & 0.5093 & 0.5044 & 0.4583 \\
    EMP6\_v6 & 0.9950 & 0.9946 & 1.0000 & 1.0000 \\
    EMP7\_v7 & 1.0000 & 1.0000 & 1.0000 & 1.0000 \\
    EMP8\_v8 & 0.1392 & 0.1442 & 1.0000 & 1.0000 \\
    GOX1\_v31\_2 & 0.0050 & 0.0131 & 0.0073 & 0.0273 \\
    GOX2\_v32 & 0.0073 & 0.0116 & 0.0079 & 0.0038 \\
    MAC1\_v37 & 0.4887 & 0.5189 & 0.4977 & 0.5106 \\
    MAC2\_v38 & 0.5217 & 0.5197 & 0.4953 & 0.5258 \\
    MGOX2\_v21 & 0.5190 & 0.4691 & 0.5196 & 0.5303 \\
    PPP10\_v18 & 0 & 0 & 0.3160 & 0.3500 \\
    PPP11\_v19 & 0 & 0 & 0.3531 & 0.3068 \\
    PPP3\_v11 & 0.4207 & 0.4706 & 1.0000 & 1.0000 \\
    PPP4\_v12 & 0.4737 & 0.4845 & 0.9997 & 0.9992 \\
    PPP5\_v13 & 0.4254 & 0.4385 & 1.0000 & 1.0000 \\
    PPP6\_v14 & 0.9970 & 0.9988 & 1.0000 & 1.0000 \\
    PPP7\_v15 & 0.3298 & 0.3438 & 1.0000 & 1.0000 \\
    PPP8\_v16 & 1.0000 & 1.0000 & 1.0000 & 1.0000 \\
    PPP9\_v17 & 0.3371 & 0.3337 & 1.0000 & 1.0000 \\
    TCA1\_v22 & 0.0010 & 0.0012 & 0.0020 & 0.0000 \\
    TCA2\_v23 & 0.3804 & 0.4172 & 0.5044 & 0.5538 \\
    TCA3\_v24 & 0.3781 & 0.3855 & 0.5061 & 0.4970 \\
    TCA4\_v25 & 0.8101 & 0.7208 & 0.0070 & 0.0114 \\
    TCA6\_v27\_2 & 0.4880 & 0.5089 & 0.8072 & 0.8871 \\
    TCA7\_v26 & 0.0027 & 0.0073 & 0.6290 & 0.7947 \\
    TCA7\_v28\_2 & 0.5143 & 0.4884 & 0.5366 & 0.6098 \\
    TCA8\_v29\_2 & 0.2525 & 0.3314 & 1.0000 & 1.0000 \\
    TCA9\_v30 & 0.4004 & 0.3747 & 1.0000 & 1.0000 \\
    TH1\_v66 & 0.4833 & 0.5043 & 0.5149 & 0.5091 \\
        \bottomrule
    \end{tabular}
    \label{tab:reproducibility}
\end{table}

\FloatBarrier
\clearpage

\section{Nomenclature}
\label{sec:nom}

\nomenclature[]{$\mathcal{D}$}{Data}

\nomenclature[]{$\mathcal{I}$, $\mathcal{I}^k$}{Finite index sets with running index $i$}
\nomenclature[]{$\mathcal{K} = \bigcup_{k=1}^{K} \mathcal{I}^k$}{Finite index set consisting of the union of $k$ index sets $\mathcal{I}^k$}

\nomenclature[]{$i^\prime, i^{\prime\prime}, k^\prime, k^{\prime\prime}$}{Sum variables}

\nomenclature[]{$\mathcal{M}$}{Single model in traditional \cmfa}
\nomenclature[]{$\mathcal{M}_i$}{Single model from model set $\{\mathcal{M}^k\}_{\mathcal{I}}$}
\nomenclature[]{$\mathcal{M}_i^k$}{Single model from model subset $\{ \mathcal{M}^k\}_{\mathcal{I}^k}$ or full model set $\{\mathcal{M}^k\}_{\mathcal{K}}$}

\nomenclature[]{$\{\mathcal{M}^k\}_{\mathcal{K}}$}{Full model set}

\nomenclature[]{$\{ \mathcal{M}^k\}_{\mathcal{I}^k}$}{Model subset}

\nomenclature[]{$\mathcal{P}_{\mathcal{M}}$, $\mathcal{P}_{\mathcal{M}^k_i}$}{Model-specific flux polytope}

\nomenclature[]{$n_x$}{Number of reactions with unknown bidirectionality in a single model $\mathcal{M}$}

\nomenclature[]{$v_{\mathcal{M}}$, $v_{\mathcal{M}_i^k}$
}{Model-specific fluxes for model $\mathcal{M}$, $\mathcal{M}_i^k$}
\nomenclature[]{$v_{\{\mathcal{M}^k\}_{\mathcal{I}^k}}$}{Model subset-averaged fluxes, averaged over all models in subset $\mathcal{M}_{\mathcal{I}^k}$}
\nomenclature[]{$v_{\{\mathcal{M}^k\}_{\mathcal{K}}}$}{Model set-averaged fluxes, averaged over all models in $\{\mathcal{M}^k\}_{\mathcal{K}}$}

\nomenclature{$p(v_{\mathcal{M}} \mid \mathcal{D})$, $p(v_{\mathcal{M}_i^k} \mid \mathcal{D})$}{Model-specific flux posterior}

\nomenclature[]{$p(\mathcal{M}_i^k \mid \{ \mathcal{M}^k\}_{\mathcal{K}}, \mathcal{D})$}{Single-model posterior probabilities in view of the full model set $\{\mathcal{M}^k\}_{\mathcal{K}}$}

\nomenclature[]{$p(\mathcal{M}_i^k \mid \{ \mathcal{M}^k\}_{\mathcal{I}^k}, \mathcal{D})$}{Single-model posterior probabilities in view of the model subset $\{\mathcal{M}^k\}_{\mathcal{I}^k}$}

\nomenclature[]{$p(\mathcal{M}_i^k)$}{Model prior in view of the full model set $\{\mathcal{M}^k\}_{\mathcal{K}}$}

\nomenclature[]{$p(\mathcal{D} \mid \mathcal{M}_i^k)$}{Model evidence of single model $\mathcal{M}_i^k$}

\nomenclature[]{$p(\mathcal{D} \mid v_{\mathcal{M}_i^k)}$}{Model-specific likelihood}

\nomenclature[]{$p(v_{\mathcal{M}_i^k} \mid \mathcal{M}_i^k)$}{Model-specific flux prior}

\nomenclature[]{$p(\mathcal{M}_i^k \mid \{ \mathcal{M}^k\}_{\mathcal{K}})$}{Model prior in view of the full model set $\{\mathcal{M}^k\}_{\mathcal{K}}$}

\nomenclature[]{$p(\mathcal{M}_i^k \mid \{ \mathcal{M}^k\}_{\mathcal{I}^k})$}{Model prior in view of the model subset $\{ \mathcal{M}^k\}_{\mathcal{I}^k}$}

\nomenclature[]{$p(v_{\{\mathcal{M}^k\}_{\mathcal{K}}} \mid \mathcal{D})$}{Model set-averaged flux posterior, averaged over the full model set $\{ \mathcal{M}^k\}_{\mathcal{K}}$}

\nomenclature[]{$p(\{ \mathcal{M}^k\}_{\mathcal{I}^k} \mid \mathcal{D})$}{Model subset posterior probability}

\nomenclature[]{$p(\mathcal{D} \mid \{ \mathcal{M}^k\}_{\mathcal{I}^k})$}{Aggregated model evidence of all models in $\{ \mathcal{M}^k \}_{\mathcal{I}^k}$}

\nomenclature[]{$p(\{ \mathcal{M}^k\}_{\mathcal{I}^k})$}{Prior of model subset $\{\mathcal{M}^k\}_{\mathcal{I}^k}$}

\nomenclature[]{$p(\mathcal{M}_i^k \mid { \{\mathcal{M}^k}_{\mathcal{I}^k}\})$}{Model prior in view of the model subset ${ \{\mathcal{M}^k\}}_{\mathcal{I}^k}$}

\printnomenclature

\clearpage
\begingroup{}
\renewcommand\refname{Supplementary References}
\bibliography{reference}

\begin{thebibliography}{41}
\providecommand{\natexlab}[1]{#1}
\providecommand{\url}[1]{\texttt{#1}}
\expandafter\ifx\csname urlstyle\endcsname\relax
  \providecommand{\doi}[1]{doi: #1}\else
  \providecommand{\doi}{doi: \begingroup \urlstyle{rm}\Url}\fi

\bibitem[Antoniewicz(2018)]{Antoniewicz2018}
M.~R. Antoniewicz.
\newblock {A guide to \textsuperscript{13}C metabolic flux analysis for the cancer biologist}.
\newblock \emph{Experimental \& molecular medicine}, 50\penalty0 (4):\penalty0 19, 2018.
\newblock \doi{10.1038/s12276-018-0060-y}.

\bibitem[Ashton et~al.(2022)Ashton, Bernstein, Buchner, Chen, Cs{\'{a}}nyi, Fowlie, Feroz, Griffiths, Handley, Habeck, Higson, Hobson, Lasenby, Parkinson, P{\'{a}}rtay, Pitkin, Schneider, Speagle, South, Veitch, Wacker, Wales, and Yallup]{Ashton2022}
G.~Ashton, N.~Bernstein, J.~Buchner, X.~Chen, G.~Cs{\'{a}}nyi, A.~Fowlie, F.~Feroz, M.~Griffiths, W.~Handley, M.~Habeck, E.~Higson, M.~Hobson, A.~Lasenby, D.~Parkinson, L.~B. P{\'{a}}rtay, M.~Pitkin, D.~Schneider, J.~S. Speagle, L.~South, J.~Veitch, P.~Wacker, D.~J. Wales, and D.~Yallup.
\newblock {Nested sampling for physical scientists}.
\newblock \emph{Nature Reviews Methods Primers}, 2\penalty0 (1):\penalty0 39, 2022.
\newblock \doi{10.1038/s43586-022-00121-x}.

\bibitem[Beste et~al.(2013)Beste, N{\"{o}}h, Niedenf{\"{u}}hr, Mendum, Hawkins, Ward, Beale, Wiechert, and McFadden]{Beste2013}
D.~J. Beste, K.~N{\"{o}}h, S.~Niedenf{\"{u}}hr, T.~A. Mendum, N.~D. Hawkins, J.~L. Ward, M.~H. Beale, W.~Wiechert, and J.~McFadden.
\newblock \textsuperscript{13}{C}-{F}lux spectral analysis of host-pathogen metabolism reveals a mixed diet for intracellular \textit{Mycobacterium tuberculosis}.
\newblock \emph{Chemistry \& Biology}, 20\penalty0 (8):\penalty0 1012--1021, 2013.
\newblock \doi{10.1016/j.chembiol.2013.06.012}.

\bibitem[{Borah Slater} et~al.(2023){Borah Slater}, Bey{\ss}, Xu, Barber, Costa, Newcombe, Theorell, Bailey, Beste, McFadden, and N{\"{o}}h]{BorahSlater2023}
K.~{Borah Slater}, M.~Bey{\ss}, Y.~Xu, J.~Barber, C.~Costa, J.~Newcombe, A.~Theorell, M.~J. Bailey, D.~J.~V. Beste, J.~McFadden, and K.~N{\"{o}}h.
\newblock One‐shot \textsuperscript{13}{C}\textsuperscript{15}{N}‐metabolic flux analysis for simultaneous quantification of carbon and nitrogen flux.
\newblock \emph{Molecular Systems Biology}, 19\penalty0 (3), 2023.
\newblock \doi{10.15252/msb.202211099}.

\bibitem[Brewer(2015)]{brewer2015inferencetransdimensionalbayesianmodels}
B.~J. Brewer.
\newblock Inference for trans-dimensional {B}ayesian models with diffusive nested sampling.
\newblock \emph{arXiv [stat.CO]]}, 1411.3921, 2015.
\newblock \doi{https://arxiv.org/abs/1411.3921}.

\bibitem[Brewer and Foreman-Mackey(2018)]{brewer2016dnest4}
B.~J. Brewer and D.~Foreman-Mackey.
\newblock {DNest4}: Diffusive nested sampling in {C++} and {P}ython.
\newblock \emph{Journal of Statistical Software}, 86\penalty0 (7):\penalty0 1--33, 2018.
\newblock \doi{10.18637/jss.v086.i07}.

\bibitem[Brewer et~al.(2011)Brewer, P{\'a}rtay, and Cs{\'a}nyi]{Brewer2011-kq}
B.~J. Brewer, L.~B. P{\'a}rtay, and G.~Cs{\'a}nyi.
\newblock Diffusive nested sampling.
\newblock \emph{Statistics and Computing}, 21\penalty0 (4):\penalty0 649--656, 2011.

\bibitem[Brewer et~al.(2015)Brewer, Huijser, and Lewis]{realTDNS}
B.~J. Brewer, D.~Huijser, and G.~F. Lewis.
\newblock {Trans-dimensional Bayesian inference for gravitational lens substructures}.
\newblock \emph{Monthly Notices of the Royal Astronomical Society}, 455\penalty0 (2):\penalty0 1819--1829, 2015.
\newblock \doi{10.1093/mnras/stv2370}.

\bibitem[Fong et~al.(2006)Fong, Nanchen, Palsson, and Sauer]{Fong2006}
S.~S. Fong, A.~Nanchen, B.~O. Palsson, and U.~Sauer.
\newblock Latent pathway activation and increased pathway capacity enable \textit{Escherichia coli} adaptation to loss of key metabolic enzymes.
\newblock \emph{Journal of Biological Chemistry}, 281:\penalty0 8024--8033, 2006.
\newblock \doi{10.1074/jbc.M510016200}.

\bibitem[Fowlie et~al.(2021)Fowlie, Handley, and Su]{fowlie2021nested}
A.~Fowlie, W.~Handley, and L.~Su.
\newblock Nested sampling with plateaus.
\newblock \emph{Monthly Notices of the Royal Astronomical Society}, 503\penalty0 (1):\penalty0 1199--1205, 2021.
\newblock \doi{10.1093/mnras/stab590}.

\bibitem[Gong et~al.(2024)Gong, Chen, Jiao, Gong, Pan, Liu, Zhang, and Tan]{Gong2024}
Z.~Gong, J.~Chen, X.~Jiao, H.~Gong, D.~Pan, L.~Liu, Y.~Zhang, and T.~Tan.
\newblock {Genome-scale metabolic network models for industrial microorganisms metabolic engineering: Current advances and future prospects}.
\newblock \emph{Biotechnology Advances}, 72:\penalty0 108319, 2024.
\newblock \doi{10.1016/j.biotechadv.2024.108319}.

\bibitem[Green(1995)]{Green1995}
P.~J. Green.
\newblock {Reversible Jump Markov Chain Monte Carlo computation and Bayesian model determination}.
\newblock \emph{Biometrika}, 82\penalty0 (4):\penalty0 711--732, 1995.

\bibitem[Hoeting et~al.(1999)Hoeting, Madigan, Raftery, and Volinsky]{Hoeting1999}
J.~A. Hoeting, D.~Madigan, A.~E. Raftery, and C.~T. Volinsky.
\newblock {Bayesian model averaging: a tutorial}.
\newblock \emph{Statistical Science}, 14\penalty0 (4):\penalty0 382 -- 417, 1999.
\newblock \doi{10.1214/ss/1009212519}.

\bibitem[Hu et~al.(2024)Hu, Baryshnikov, and Handley]{Hu2024}
Z.~Hu, A.~Baryshnikov, and W.~Handley.
\newblock {AEONS: approximating the end of nested sampling}.
\newblock \emph{Monthly Notices of the Royal Astronomical Society}, 532\penalty0 (4):\penalty0 4035--4049, 2024.
\newblock \doi{10.1093/mnras/stae1754}.

\bibitem[Jadebeck et~al.(2020)Jadebeck, Theorell, Leweke, and Nöh]{HOPS}
J.~F. Jadebeck, A.~Theorell, S.~Leweke, and K.~Nöh.
\newblock {HOPS}: high-performance library for (non-)uniform sampling of convex-constrained models.
\newblock \emph{Bioinformatics}, 37\penalty0 (12):\penalty0 1776--1777, 2020.
\newblock \doi{10.1093/bioinformatics/btaa872}.

\bibitem[Jadebeck et~al.(2023)Jadebeck, Wiechert, and N{\"o}h]{Jadebeck2023}
J.~F. Jadebeck, W.~Wiechert, and K.~N{\"o}h.
\newblock Practical sampling of constraint-based models: Optimized thinning boosts {CHRR} performance.
\newblock \emph{PLOS Computational Biology}, 19\penalty0 (8):\penalty0 e1011378, 2023.

\bibitem[Jadebeck et~al.(2025)Jadebeck, Wiechert, and Nöh]{jadebeck2025}
J.~F. Jadebeck, W.~Wiechert, and K.~Nöh.
\newblock Trans-dimensional diffusive nested sampling for metabolic network inference.
\newblock \emph{Physical Sciences Forum}, 12\penalty0 (1):\penalty0 5, 2025.
\newblock \doi{10.3390/psf2025012005}.

\bibitem[Kappelmann et~al.(2016)Kappelmann, Wiechert, and Noack]{Kappelmann2016}
J.~Kappelmann, W.~Wiechert, and S.~Noack.
\newblock {Cutting the Gordian Knot: Identifiability of anaplerotic reactions in \textit{Corynebacterium glutamicum} by means of \textsuperscript{13}C-metabolic flux analysis}.
\newblock \emph{Biotechnology and Bioengineering}, 113\penalty0 (3):\penalty0 661--674, 2016.
\newblock \doi{10.1002/bit.25833}.

\bibitem[Kass and Raftery(1995)]{kass1995bayes}
R.~E. Kass and A.~E. Raftery.
\newblock Bayes factors.
\newblock \emph{Journal of the American Statistical Association}, 90\penalty0 (430):\penalty0 773--795, 1995.
\newblock \doi{10.1080/01621459.1995.10476572}.

\bibitem[Leighty and Antoniewicz(2013)]{Leighty2013}
R.~W. Leighty and M.~R. Antoniewicz.
\newblock {COMPLETE-MFA: Complementary parallel labeling experiments technique for metabolic flux analysis}.
\newblock \emph{Metabolic Engineering}, 20:\penalty0 49--55, 2013.
\newblock \doi{10.1016/j.ymben.2013.08.006}.

\bibitem[Linden-Santangeli and Rangamani(2025)]{Linden-Santangeli2025}
N.~Linden-Santangeli and P.~Rangamani.
\newblock {Increasing certainty in systems biology models using Bayesian multimodel inference}.
\newblock \emph{Nature Communications}, 16:\penalty0 7416, 2025.
\newblock \doi{10.1038/s41467-025-62415-4}.

\bibitem[Liu et~al.(2025)Liu, Ding, Wang, Ren, Lee, and Zhang]{Liu2025}
L.~Liu, D.~Ding, H.~Wang, X.~Ren, S.~Y. Lee, and D.~Zhang.
\newblock {Balancing cell growth and product synthesis for efficient microbial cell factories}.
\newblock \emph{Advanced Science}, 12\penalty0 (40), 2025.
\newblock \doi{10.1002/advs.202510649}.

\bibitem[Long and Antoniewicz(2019)]{Long2019}
C.~P. Long and M.~R. Antoniewicz.
\newblock {High-resolution \textsuperscript{13}C metabolic flux analysis}.
\newblock \emph{Nature Protocols}, 14\penalty0 (10):\penalty0 2856--2877, 2019.
\newblock \doi{10.1038/s41596-019-0204-0}.

\bibitem[MacKay(2008)]{MacKay2008}
D.~J. MacKay.
\newblock \emph{{Information Theory, Inference, and Learning Algorithms}}.
\newblock Cambridge University Press, Cambridge, 2008.
\newblock ISBN 0521642981.
\newblock \doi{10.2277/0521642981}.

\bibitem[McFadden(2023)]{Johnjoe2023}
J.~McFadden.
\newblock {Razor sharp: The role of Occam's razor in science}.
\newblock \emph{Annals of the New York Academy of Sciences}, 1530\penalty0 (1):\penalty0 8--17, 2023.
\newblock \doi{https://doi.org/10.1111/nyas.15086}.

\bibitem[Niedenf{\"{u}}hr et~al.(2015)Niedenf{\"{u}}hr, Wiechert, and N{\"{o}}h]{Niedenfuhr2015}
S.~Niedenf{\"{u}}hr, W.~Wiechert, and K.~N{\"{o}}h.
\newblock {How to measure metabolic fluxes: A taxonomic guide for \textsuperscript{13}C fluxomics}.
\newblock \emph{Current Opinion in Biotechnology}, 34:\penalty0 82--90, 2015.
\newblock \doi{10.1016/j.copbio.2014.12.003}.

\bibitem[Nishikawa et~al.(2008)Nishikawa, Gulbahce, and Motter]{Nishikawa2008}
T.~Nishikawa, N.~Gulbahce, and A.~E. Motter.
\newblock {Spontaneous reaction silencing in metabolic optimization}.
\newblock \emph{PLoS computational biology}, 4\penalty0 (12):\penalty0 e1000236, 2008.
\newblock \doi{10.1371/journal.pcbi.1000236}.

\bibitem[Paul et~al.(2024)Paul, Jadebeck, Stratmann, Wiechert, and Nöh]{hopsy}
R.~D. Paul, J.~F. Jadebeck, A.~Stratmann, W.~Wiechert, and K.~Nöh.
\newblock hopsy - a methods marketplace for convex polytope sampling in python.
\newblock \emph{Bioinformatics}, 40\penalty0 (7):\penalty0 btae430, 2024.
\newblock \doi{10.1093/bioinformatics/btae430}.

\bibitem[Sauer(2006)]{Sauer2006}
U.~Sauer.
\newblock Metabolic networks in motion: \textsuperscript{13}{C}-based flux analysis.
\newblock \emph{Molecular Systems Biology}, 2:\penalty0 62, 2006.
\newblock \doi{10.1038/msb4100109}.

\bibitem[Schittenhelm and Wacker(2021)]{schittenhelm2021nestedsamplinglikelihoodplateaus}
D.~Schittenhelm and P.~Wacker.
\newblock Nested sampling and likelihood plateaus.
\newblock \emph{arXiv preprint arXiv:2005.08602}, 2021.

\bibitem[Skilling(2006)]{Skilling2006}
J.~Skilling.
\newblock {Nested sampling for general Bayesian computation}.
\newblock \emph{Bayesian Analysis}, 1\penalty0 (4):\penalty0 833 -- 859, 2006.
\newblock \doi{10.1214/06-BA127}.

\bibitem[Stratmann et~al.(2025)Stratmann, Bey{\ss}, Jadebeck, Wiechert, and N{\"{o}}h]{Stratmann2025}
A.~Stratmann, M.~Bey{\ss}, J.~F. Jadebeck, W.~Wiechert, and K.~N{\"{o}}h.
\newblock {13CFLUX - third-generation high-performance engine for isotopically (non)stationary 13C metabolic flux analysis}.
\newblock \emph{Bioinformatics}, 41\penalty0 (12):\penalty0 6, 2025.
\newblock \doi{10.1093/bioinformatics/btaf630}.

\bibitem[Sundqvist et~al.(2022)Sundqvist, Grankvist, Watrous, Mohit, Nilsson, and Cedersund]{Sundqvist2022}
N.~Sundqvist, N.~Grankvist, J.~Watrous, J.~Mohit, R.~Nilsson, and G.~Cedersund.
\newblock {Validation-based model selection for \textsuperscript{13}C metabolic flux analysis with uncertain measurement errors}.
\newblock \emph{PLOS Computational Biology}, 18\penalty0 (4):\penalty0 e1009999, 2022.
\newblock \doi{10.1371/journal.pcbi.1009999}.

\bibitem[Theorell et~al.(2017)Theorell, Leweke, Wiechert, and Nöh]{Theorell2017}
A.~Theorell, S.~Leweke, W.~Wiechert, and K.~Nöh.
\newblock {To be certain about the uncertainty: Bayesian statistics for \textsuperscript{13}C metabolic flux analysis}.
\newblock \emph{Biotechnology and Bioengineering}, 114\penalty0 (11):\penalty0 2668--2684, 2017.
\newblock \doi{https://doi.org/10.1002/bit.26379}.

\bibitem[Theorell et~al.(2022)Theorell, Jadebeck, N{\"{o}}h, and Stelling]{Theorell2022}
A.~Theorell, J.~F. Jadebeck, K.~N{\"{o}}h, and J.~Stelling.
\newblock {PolyRound: polytope rounding for random sampling in metabolic networks}.
\newblock \emph{Bioinformatics}, 38\penalty0 (2):\penalty0 566--567, 2022.
\newblock \doi{10.1093/bioinformatics/btab552}.

\bibitem[Theorell et~al.(2024)Theorell, Jadebeck, Wiechert, McFadden, and N{\"{o}}h]{Theorell2024}
A.~Theorell, J.~F. Jadebeck, W.~Wiechert, J.~McFadden, and K.~N{\"{o}}h.
\newblock {Rethinking \textsuperscript{13}C-metabolic flux analysis – The Bayesian way of flux inference}.
\newblock \emph{Metabolic Engineering}, 83:\penalty0 137--149, 2024.
\newblock \doi{10.1016/j.ymben.2024.03.005}.

\bibitem[Wiechert(2007)]{Wiechert2007}
W.~Wiechert.
\newblock The thermodynamic meaning of metabolic exchange fluxes.
\newblock \emph{Biophysical journal}, 93\penalty0 (6):\penalty0 2255--2264, 2007.

\bibitem[Wiechert and de~Graaf(1997)]{Wiechert1997}
W.~Wiechert and A.~A. de~Graaf.
\newblock {Bidirectional reaction steps in metabolic networks. Part I. Modeling and simulation of carbon isotope labeling experiments}.
\newblock \emph{Biotechnology and Bioengineering}, 55\penalty0 (1):\penalty0 101--117, 1997.
\newblock \doi{10.1002/(SICI)1097-0290(19970705)55:1<101::AID-BIT12>3.0.CO;2-P}.

\bibitem[Wiechert and N{\"{o}}h(2021)]{Wiechert2021}
W.~Wiechert and K.~N{\"{o}}h.
\newblock {Quantitative metabolic flux analysis based on isotope labeling}.
\newblock In J.~Nielsen, G.~Stephanopoulos, and S.~Y. Lee, editors, \emph{Metabolic Engineering: Concepts and Applications}, chapter~3, pages 73--136. Wiley, 2021.
\newblock \doi{10.1002/9783527823468.ch3}.

\bibitem[Wiechert et~al.(1997)Wiechert, Siefke, de~Graaf, and Marx]{Wiechert1997bidir2}
W.~Wiechert, C.~Siefke, A.~A. de~Graaf, and A.~Marx.
\newblock {Bidirectional reaction steps in metabolic networks: II. Flux estimation and statistical analysis}.
\newblock \emph{Biotechnology and Bioengineering}, 55\penalty0 (1):\penalty0 118–135, 1997.
\newblock \doi{10.1002/(sici)1097-0290(19970705)55:1<118::aid-bit13>3.0.co;2-i}.

\bibitem[Zamboni et~al.(2009)Zamboni, Fendt, R{\"{u}}hl, and Sauer]{Zamboni2009}
N.~Zamboni, S.-M. Fendt, M.~R{\"{u}}hl, and U.~Sauer.
\newblock {\textsuperscript{13}C-based metabolic flux analysis}.
\newblock \emph{Nature Protocols}, 4:\penalty0 878--892, 2009.
\newblock \doi{10.1038/nprot.2009.58}.

\end{thebibliography}


\begin{thebibliography}{}

\bibitem[Antoniewicz, 2018]{Antoniewicz2018}
Antoniewicz, M.R. (2018).
\newblock A guide to \textsuperscript{13}C metabolic flux analysis for the cancer biologist.
\newblock {\em Exp Mol Med}, \textbf{50}: 19.

\bibitem[Beste et~al., 2013]{Beste2013}
Beste, D.J. \emph{et al}. (2013).
\newblock \textsuperscript{13}C-flux spectral analysis of host-pathogen metabolism reveals a mixed diet for intracellular \textit{Mycobacterium tuberculosis}.
\newblock {\em Chem Biol}, \textbf{20}: 1012--1021.

\bibitem[Borah Slater et~al.(2023)Borah Slater, Beyß, Xu, Barber, Costa, Newcombe, Theorell, Bailey, Beste, McFadden, and N{\"{o}}h]{BorahSlater2023}
Borah Slater, K. \emph{et al}. (2023).
\newblock One-shot $^{13}$C$^{15}$N metabolic flux analysis for simultaneous quantification of carbon and nitrogen flux.
\newblock \emph{Mol Syst Biol}, \textbf{19}\penalty0 (3):\penalty0 e11099.

\bibitem[Brewer, 2015]{Brewer2015_inference}
Brewer, B.J. (2015).
\newblock Inference for trans-dimensional Bayesian models with diffusive nested sampling.
\newblock {\em arXiv preprint arXiv:1411.3921}.

\bibitem[Brewer and Foreman-Mackey, 2018]{Brewer2016}
Brewer, B.J. and Foreman-Mackey, D. (2018).
\newblock {DNest4}: diffusive nested sampling in {C++} and Python.
\newblock {\em J Stat Softw}, \textbf{86}: 1--33.

\bibitem[Brewer et~al., 2015]{Brewer2015_lens}
Brewer, B.J. \emph{et al}. (2015).
\newblock Trans-dimensional Bayesian inference for gravitational lens substructures.
\newblock {\em Mon Not R Astron Soc}, \textbf{455}: 1819--1829.

\bibitem[Brewer et~al., 2011]{Brewer2011}
Brewer, B.J. \emph{et al}. (2011).
\newblock Diffusive nested sampling.
\newblock {\em Stat Comput}, \textbf{21}: 649--656.

\bibitem[Fong et~al., 2006]{Fong2006}
Fong, S.S. \emph{et al}. (2006).
\newblock Latent pathway activation and increased pathway capacity enable \textit{Escherichia coli} adaptation to loss of key metabolic enzymes.
\newblock {\em J Biol Chem}, \textbf{281}: 8024--8033.

\bibitem[Gong et~al., 2024]{Gong2024}
Gong, Z. \emph{et al}. (2024).
\newblock Genome-scale metabolic network models for industrial microorganisms metabolic engineering: current advances and future prospects.
\newblock {\em Biotechnol Adv}, \textbf{72}: 108319.

\bibitem[Green, 1995]{Green1995}
Green, P.J. (1995).
\newblock Reversible jump Markov chain Monte Carlo computation and Bayesian model determination.
\newblock {\em Biometrika}, \textbf{82}: 711--732.

\bibitem[Hoeting et~al., 1999]{Hoeting1999}
Hoeting, J.A. \emph{et al}. (1999).
\newblock Bayesian model averaging: a tutorial. 
\newblock {\em Stat Sci}, \textbf{14}: 382--417.

\bibitem[Jadebeck et~al., 2020]{HOPS}
Jadebeck, J.F. \emph{et al}. (2020).
\newblock {HOPS}: high-performance library for (non-)uniform sampling of convex-constrained models.
\newblock {\em Bioinformatics}, \textbf{37}: 1776--1777.

\bibitem[Jadebeck et~al., 2023]{Jadebeck2023}
Jadebeck, J.F. \emph{et al}. (2023).
\newblock Practical sampling of constraint-based models: optimized thinning boosts {CHRR} performance.
\newblock {\em PLoS Comput Biol}, \textbf{19}: e1011378.

\bibitem[Kappelmann et~al., 2016]{Kappelmann2016}
Kappelmann, J. \emph{et al}. (2016).
\newblock Cutting the Gordian knot: identifiability of anaplerotic reactions in \textit{Corynebacterium glutamicum} by means of \textsuperscript{13}C-metabolic flux analysis.
\newblock {\em Biotechnol Bioeng}, \textbf{113}: 661--674.

\bibitem[Kass and Raftery, 1995]{kass1995bayes}
Kass, R.E. and Raftery, A.E. (1995).
\newblock Bayes factors.
\newblock {\em J Am Stat Assoc}, \textbf{90}: 773--795.

\bibitem[Leighty and Antoniewicz, 2013]{Leighty2013}
Leighty, R.W. and Antoniewicz, M.R. (2013).
\newblock {COMPLETE-MFA}: complementary parallel labeling experiments technique for metabolic flux analysis.
\newblock {\em Metab Eng}, \textbf{20}: 49--55.

\bibitem[Linden-Santangeli and Rangamani, 2025]{Linden-Santangeli2025}
Linden-Santangeli, N. and Rangamani, P. (2025).
\newblock {Increasing certainty in systems biology models using Bayesian multimodel inference}.
\newblock {\em Nat Com}, \textbf{16}: 7416.

\bibitem[Liu et~al., 2025]{Liu2025}
Liu, L. \emph{et al}. (2025).
\newblock Balancing cell growth and product synthesis for efficient microbial cell factories.
\newblock {\em Adv Sci}, \textbf{12}: e10649.

\bibitem[Long and Antoniewicz, 2019]{Long2019}
Long, C.P. and Antoniewicz, M.R. (2019).
\newblock High-resolution \textsuperscript{13}C metabolic flux analysis.
\newblock {\em Nat Protoc}, \textbf{14}: 2856--2877.

\bibitem[MacKay, 2008]{MacKay2008}
MacKay, D.J.C. (2008).
\newblock {\em Information Theory, Inference, and Learning Algorithms}.
\newblock Cambridge University Press, Cambridge.

\bibitem[McFadden, 2023]{Johnjoe2023}
McFadden, J. (2023).
\newblock Razor sharp: the role of Occam's razor in science.
\newblock {\em Ann N Y Acad Sci}, \textbf{1530}: 8--17.


\bibitem[Niedenf{\"u}hr et~al., 2015]{Niedenfuhr2015}
Niedenf{\"u}hr, S. \emph{et al}. (2015).
\newblock How to measure metabolic fluxes: a taxonomic guide for \textsuperscript{13}C fluxomics.
\newblock {\em Curr Opin Biotechnol}, \textbf{34}: 82--90.

\bibitem[Nishikawa et~al., 2008]{Nishikawa2008}
Nishikawa, T. \emph{et al}. (2008).
\newblock Spontaneous reaction silencing in metabolic optimization.
\newblock {\em PLoS Comput Biol}, \textbf{4}: e1000236.

\bibitem[Paul et~al., 2024]{hopsy}
Paul, R.D. \emph{et al}. (2024).
\newblock {hopsy}: a methods marketplace for convex polytope sampling in Python.
\newblock {\em Bioinformatics}, \textbf{40}: btae430.

\bibitem[Sauer, 2006]{Sauer2006}
Sauer, U. (2006).
\newblock Metabolic networks in motion: \textsuperscript{13}C-based flux analysis.
\newblock {\em Mol Syst Biol}, \textbf{2}: 62.


\bibitem[Stratmann et~al., 2025]{Stratmann2025}
Stratmann, A. \emph{et al}. (2025).
\newblock {13CFLUX} -- third-generation high-performance engine for isotopically (non)stationary \textsuperscript{13}C metabolic flux analysis.
\newblock {\em arXiv preprint arXiv:2509.23847}.

\bibitem[Sundqvist et~al., 2022]{Sundqvist2022}
Sundqvist, N. \emph{et al}. (2022).
\newblock Validation-based model selection for \textsuperscript{13}C metabolic flux analysis with uncertain measurement errors.
\newblock {\em PLoS Comput Biol}, \textbf{18}: e1009999.

\bibitem[Theorell et~al., 2022]{Theorell2022}
Theorell, A. \emph{et al}. (2022).
\newblock {PolyRound}: polytope rounding for random sampling in metabolic networks.
\newblock {\em Bioinformatics}, \textbf{38}: 566--567.

\bibitem[Theorell et~al., 2024]{Theorell2024}
Theorell, A. \emph{et al}. (2024).
\newblock Rethinking \textsuperscript{13}C metabolic flux analysis -- the Bayesian way of flux inference.
\newblock {\em Metab Eng}, \textbf{83}: 137--149.

\bibitem[Theorell et~al., 2017]{Theorell2017}
Theorell, A. \emph{et al}. (2017).
\newblock To be certain about the uncertainty: Bayesian statistics for \textsuperscript{13}C metabolic flux analysis.
\newblock {\em Biotechnol Bioeng}, \textbf{114}: 2668--2684.

\bibitem[Theorell and N{\"o}h, 2020]{Theorell2020}
Theorell, A. and N{\"o}h, K. (2020).
\newblock Reversible jump {MCMC} for multi-model inference in metabolic flux analysis.
\newblock {\em Bioinformatics}, \textbf{36}: 232--240.

\bibitem[Wiechert, 2007]{Wiechert2007}
Wiechert, W. (2007).
\newblock The thermodynamic meaning of metabolic exchange fluxes.
\newblock {\em Biophys J}, \textbf{93}: 2255--2264.

\bibitem[Wiechert and de~Graaf, 1997]{Wiechert1997}
Wiechert, W. and de~Graaf, A.A. (1997).
\newblock Bidirectional reaction steps in metabolic networks. Part I. Modeling and simulation of carbon isotope labeling experiments.
\newblock {\em Biotechnol Bioeng}, \textbf{55}: 101--117.

\bibitem[Wiechert and N{\"o}h, 2021]{Wiechert2021}
Wiechert, W. and N{\"o}h, K. (2021).
\newblock Quantitative metabolic flux analysis based on isotope labeling.
\newblock In: Nielsen, J., Stephanopoulos, G. and Lee, S.-Y. (eds), {\em Metabolic Engineering: Concepts and Applications}.
\newblock Wiley-VCH, Weinheim, Germany, pp.\ 73--136.

\bibitem[Wiechert et~al., 1997]{Wiechert1997bidir2}
Wiechert, W. \emph{et al}. (1997).
\newblock Bidirectional reaction steps in metabolic networks: II. Flux estimation and statistical analysis.
\newblock {\em Biotechnol Bioeng}, \textbf{55}: 118--135.

\bibitem[Zamboni et~al., 2009]{Zamboni2009}
Zamboni, N. \emph{et al}. (2009).
\newblock \textsuperscript{13}C-based metabolic flux analysis.
\newblock {\em Nat Protoc}, \textbf{4}: 878--892.

\end{thebibliography}
\endgroup{}
\end{document}

\typeout{get arXiv to do 4 passes: Label(s) may have changed. Rerun}